\documentclass[aps,prd,twocolumn,amsmath,amssymb，preprintnumbers,superscriptaddress]{revtex4-2}

\usepackage{fontspec}    
\usepackage{xeCJK}   
\usepackage[normalem]{ulem} 

\usepackage{graphicx}
\usepackage{dcolumn}
\usepackage{bm}
\usepackage{multirow}
\usepackage{comment}
\usepackage{upgreek}

\def\mglong{\texttt{MadGraph5\_aMC@NLO}}
\def\mgshort{\texttt{MG5\_aMC}}
\def\pythia{\texttt{Pythia8.3}}

 \usepackage{ifpdf}
 \ifpdf
 \usepackage[pdftex]{hyperref}
 \else
 \usepackage{hyperref}
 \fi

 \hypersetup{
   pdftitle={},
   pdfauthor={},
   pdfsubject={},
   pdfkeywords={},
   pdfstartview={},
   bookmarksopen=true, breaklinks=true, debug=true, 
   colorlinks=true, linkcolor=red, citecolor=blue, urlcolor=blue
 }

\begin{document}

\title{Prospects for discovering strongly decaying doubly heavy $T_{bc}$ tetraquark states at LHCb}

\thanks{The work of MF and YL is supported by the National Natural Science Foundation of China (NSFC) (grant No.12175245, 12188102, W2443008). 
The work of HSS is supported by the European Research Council (grant No.101041109, ``BOSON'') and the French National Research Agency  (grant ANR-20-CE31-0015, ``PrecisOnium'').}

\author{Mingjie Feng (冯铭婕)}
\affiliation{Institute of High Energy Physics, Chinese Academy of Science, Beijing 100049, China}
\affiliation{University of Chinese Academy of Science, Beijing 100049, China}

\author{Yiming Li (李一鸣)}
\email{liyiming@ihep.ac.cn}
\affiliation{Institute of High Energy Physics, Chinese Academy of Science, Beijing 100049, China}

\author{Hua-Sheng Shao (邵华圣)}
\affiliation{Laboratoire de Physique Th\'eorique et Hautes Energies (LPTHE), UMR 7589,
Sorbonne Universit\'e et CNRS, 4 place Jussieu, 75252 Paris, France}
                 
\date{\today}

\begin{abstract}
We investigate the discovery potential of the $T_{bc}$ state with $J^P = 0^+$ in proton-proton ($pp$) collisions at LHCb at a center-of-mass energy of $\sqrt{s} = 13~\mathrm{TeV}$. The study focuses on the decay channel $T_{bc} \to B^- D^+$. A phenomenological approach is employed to construct the background model based on the associated production of $B$ and $D$ mesons, incorporating previously published LHCb results. Background processes are simulated using \mglong\ and \pythia. We explore the parameter space of the $T_{bc}$ mass, width, production cross section, and the effective double-parton scattering cross section ($\sigma_{\mathrm{eff}}$) relevant for the $B D$ meson background. The integrated luminosity required for a $5\sigma$ discovery at LHCb is evaluated under various assumptions. In particular, we consider three representative $T_{bc}$ production cross section scenarios: an optimistic estimate of $103~\mathrm{nb}$, an intermediate value of $18~\mathrm{nb}$ obtained by scaling from the $T_{cc}^+$ production cross section, and a conservative lower bound of $0.3~\mathrm{nb}$. We find that a $5\sigma$ observation is achievable for a production cross section of $103~\mathrm{nb}$, which is expected to be within reach during Run~4. In contrast, the more realistic cross section estimate of $18~\mathrm{nb}$ requires the full Run~5 dataset ($300~\mathrm{fb}^{-1}$) under the most favorable parameter choices. For the conservative scenario, no significant signal would be observable even with $300~\mathrm{fb}^{-1}$. In addition, we estimate the minimum observable $\sigma(T_{bc}) \times \mathcal{B}(T_{bc} \to B^- D^+)$ for a $5\sigma$ discovery under different luminosity scenarios, providing guidance for future experimental searches at LHCb.
\end{abstract}


\maketitle

\section{\label{sec:intro}Introduction}

Since the discovery of the $X(3872)$ by the Belle experiment in 2003~\cite{Belle:2003nnu}, numerous resonant states have been observed that lie beyond the conventional quark-model description of the hadron spectrum. These observations have significantly broadened our understanding of hadron spectroscopy. Exotic hadrons also play an important role in the study of quantum chromodynamics (QCD), as they provide a unique window into the non-perturbative dynamics of the strong interaction. Consequently, the investigation of exotic hadronic states has become a frontier of hadron physics over the past two decades. A wide variety of theoretical models have been proposed, accompanied by extensive experimental efforts to search for new exotic states via various production mechanisms and to measure their masses, widths, production cross sections, and spin-parity quantum numbers to elucidate their internal structure.

The observation of the doubly charmed tetraquark $T_{cc}^+$ and the doubly charmed baryon $\Xi_{cc}^{++}$ in proton–proton collisions~\cite{LHCb:2021vvq,LHCb:2018pcs} has firmly established that hadrons containing two heavy quarks can be produced promptly at hadron colliders. These discoveries open a new frontier in the search for other doubly heavy systems, among which the $T_{bc}$ and $T_{bb}$ tetraquarks have yet to be observed experimentally.

Unlike the doubly charmed case, the $T_{bc}$ contains two different heavy flavors, namely a bottom quark and a charm quark. This distinctive composition makes it an invaluable probe of non-perturbative QCD dynamics in systems with unequal heavy-quark masses. It provides a unique opportunity to test heavy-quark symmetry and its breaking, explore the flavor dependence of diquark correlations, and constrain models of multiquark binding~\cite{Deng:2021gnb, Cheng:2020wxa}. Its discovery would enrich the spectrum of doubly heavy tetraquarks and deepen our understanding of exotic hadron structure.

Numerous theoretical predictions exist for the ground-state mass of the $T_{bc}(bc\overline{u}\overline{d})$ tetraquark (charge conjugation is implied throughout this paper). The $T_{bc}$ mass is generally expected to lie close to the open-bottom–open-charm meson threshold. The predicted mass difference with respect to the $\bar{B} D$ threshold falls in the range~\cite{Chen:2013aba, Karliner:2017qjm, Carames:2018tpe, Radhakrishnan:2024ihu, Meinel:2022lzo, Wu:2024zbx, Ebert:2007rn, Eichten:2017ffp, Cheng:2020wxa, Song:2023izj}
\begin{equation}
-50~\mathrm{MeV} < m(T_{bc})-m(\bar{B})-m(D) < 120~\mathrm{MeV}\,.
\end{equation}
At present, it remains unclear whether the $T_{bc}$ mass lies above or below the $\bar{B}D$ threshold. If the mass lies below threshold, the $T_{bc}$ would be stable against strong decays and could only be accessed via weak interactions~\cite{Chen:2013aba, Karliner:2017qjm, Carames:2018tpe, Radhakrishnan:2024ihu, Meinel:2022lzo, Wu:2024zbx}. Detailed studies of the weak decay modes of such a doubly heavy tetraquark, including both semileptonic and nonleptonic channels induced by $b$- or $c$-quark decays, have been performed using flavor SU(3) symmetry in ref.~\cite{Xing:2018bqt}. It is worth noting that weak decay channels may also provide promising discovery opportunities for $T_{bc}$ states at LHCb, and related experimental studies are ongoing. If the mass lies above threshold, the $T_{bc}$ is expected to appear as a relatively narrow resonance in the $\bar{B}D$ invariant-mass spectrum and can therefore be searched for directly~\cite{Ebert:2007rn, Eichten:2017ffp, Cheng:2020wxa, Song:2023izj}. In this study, we focus on the latter scenario, where the $T_{bc}$ lies above the threshold and can be identified through its narrow resonance in the $\bar{B}D$ invariant-mass spectrum. Studies of heavy-flavor hadron-associated production thus provide a powerful experimental avenue for the discovery of new hadronic states containing different heavy quarks.

A central experimental challenge in searches for doubly heavy tetraquarks at hadron colliders is the quantitative understanding of their prompt production and the associated backgrounds. In particular, the prompt production of heavy-meson pairs, such as $BD$, forms an irreducible background for tetraquark states decaying into two heavy mesons. A realistic description of such backgrounds is therefore essential for evaluating the experimental sensitivity and discovery potential for new states like $T_{bc}$.

From the perspective of production mechanisms, associated $BD$ production proceeds primarily through single parton scattering (SPS) and multiparton scattering. The latter is dominated by double parton scattering (DPS), while contributions from higher-order multiparton interactions are expected to be suppressed. The relative importance of DPS is expected to increase with increasing proton-proton collision energy. Studies of associated $BD$ production provide a sensitive probe of heavy-flavor production mechanisms, enable quantitative tests of SPS and DPS dynamics (see, e.g., ref.~\cite{Maciula:2018mig}), constrain heavy-quark correlations and fragmentation, and supply essential background information for searches for exotic states such as the $T_{bc}$ tetraquark.

Previous theoretical studies have primarily focused on predicting the $T_{bc}$ mass and width. However, a critical gap remains: a quantitative, data-driven assessment of the actual discovery potential at LHCb. This work aims to fill this gap by addressing the question: under what specific conditions -- in terms of integrated luminosity, production cross section, and branching fraction -- can a $5\sigma$ discovery of the $T_{bc}$ be achieved amidst the formidable $BD$ background? We provide a realistic evaluation of these conditions based on simulations calibrated to LHCb data.

Owing to the high heavy-flavor production rate in $pp$ collisions, its excellent vertexing and tracking capability, efficient trigger, and precise control of detector-induced asymmetries, the LHCb experiment provides an ideal environment for searches for the $T_{bc}$.

The $T_{bc}(bc\overline{u}\overline{d})$ tetraquark with quantum numbers $J^P = 0^+$ are expected to decay strongly into $\bar{B}^0 D^0$, $B^- D^+$, and their charge-conjugate modes if the mass is above the meson-pair thresholds. This analysis investigates the experimental potential of the LHCb experiment to discover the $T_{bc}$ through its production and decay into $BD$ meson pairs. The study focuses on the decay chain
\begin{equation}
\begin{split}
T_{bc} &\to B^-D^+\,, \\
B^- &\to J/\psi(\to \mu^+\mu^-) K^-\,, \\
D^+ &\to K^-\pi^+\pi^+\,.
\label{eq:Tbcdecaychain}
\end{split}
\end{equation}
This provides a clean experimental signature for the discovery of the $T_{bc}$ at LHCb.

Our approach maintains a close connection to experimental reality. The DPS background is calibrated using LHCb data, while the SPS contribution is evaluated with two different flavor-scheme calculations to account for theoretical uncertainties. Rather than relying on a single theoretical prediction, we treat $\sigma(T_{bc}) \times \mathcal{B}(T_{bc} \to B^-D^+)$ as a free parameter and scan over a well-motivated range, together with the $T_{bc}$ mass, width, and effective DPS cross section $\sigma_{\mathrm{eff}}$. After incorporating realistic reconstruction efficiencies, we translate the signal and background yields into the integrated luminosity required for a $5\sigma$ discovery. This work provides a quantitative benchmark for the $T_{bc}$ search at LHCb and serves as a reference for future experimental studies in background-dominated environments.

\section{\label{sec:method}Method}

\subsection{Background Simulations}
\label{sec:bkg}

In this analysis, both SPS and DPS processes are considered the dominant mechanisms contributing to the background in the $B^\mp D^\pm$ final state in the search for $T_{bc}$. The background samples for the two processes are generated separately, with the DPS sample reweighted to reproduce the differential distributions observed by LHCb. The following kinematic selection cuts are applied to both background samples to reflect the acceptance of the LHCb detector: 
\begin{equation}
\begin{split}
2.0 < y_{B/D} &< 4.5\,, \\
p_{\mathrm{T},B} &< 40~\mathrm{GeV}\,, \\
1~\mathrm{GeV} < p_{\mathrm{T},D} &< 8~\mathrm{GeV}\,.
\label{eq:FDcuts}
\end{split}
\end{equation}
The SPS and DPS background samples are generated using \mglong\ (\mgshort\ hereafter)~\cite{Alwall:2014hca,Frederix:2018nkq}, with parton showering and hadronization performed by \pythia~\cite{Bierlich:2022pfr}. We consider proton-proton collisions at $\sqrt{s}=13$ TeV.
The charm- and bottom-quark masses are set to $m_c = 1.55~\mathrm{GeV}$ and $m_b = 4.7~\mathrm{GeV}$, respectively. 
The \texttt{NNPDF2.3} parton distribution functions (PDFs)~\cite{Ball:2012cx} are employed, with next-to-leading order (NLO) PDFs used for NLO simulations and leading order (LO) PDFs for LO simulations. Both the renormalization and factorization scales are chosen dynamically, with the central value given by $\mu_0=H_T/2$, where $H_T$ is the scalar sum of the transverse energies of the final-state particles. While the scale choice for the background process $pp \to b\bar{b}c\bar{c}+X$ is not uniquely defined, using $H_T/2$ is common practice in modern event generators, as it provides a natural estimate of the typical momentum transfer in such processes.

For simulating the SPS background, we consider two complementary approaches to generate samples. First, we study the process $pp \to b\bar{b}c\bar{c}+X$ in the $3$-flavor number scheme (3FNS) at NLO in QCD, with matching to parton showers performed using the \texttt{MC@NLO} method~\cite{Frixione:2002ik}. We take the 3FNS calculation as the nominal prediction. The corresponding fiducial cross section, with fiducial cuts given in eq.~\eqref{eq:FDcuts}, is found to be
\begin{equation}
\sigma^{\mathrm{NLO~SPS~3FNS}}_{B^\mp D^\pm} = 0.074^{+0.040}_{-0.034}~\upmu\mathrm{b}\,.
\end{equation}
The uncertainty is dominated by variations in the renormalization and factorization scales. The fiducial cross section includes both $B^-D^+$ and $B^+D^-$ final states. Together with the DPS contribution, this NLO SPS prediction defines the \textit{baseline background scenario} used to assess the $T_{bc}$ discovery potential.

Alternatively, as discussed in ref.~\cite{Shao:2020kgj}, the resummation of initial-state logarithms generated by gluon splitting into a charm-quark pair may be crucial for describing the LHCb $J/\psi+D$ measurement~\cite{LHCb:2012aiv}. We therefore also generate the SPS sample at LO with charm- or anticharm-gluon initial states in the $4$-flavor number scheme (4FNS). This LO 4FNS calculation yields a fiducial cross section of
\begin{equation}
\sigma_{B^\mp D^\pm}^{\mathrm{LO~SPS~4FNS}} = 0.21^{+0.44}_{-0.15}~\upmu\mathrm{b}\,.
\end{equation}
The dominant theoretical uncertainty arises from scale variations. As in the $J/\psi+D$ case~\cite{Shao:2020kgj}, the 4FNS calculation significantly increases the SPS fiducial cross section relative to the 3FNS result. To obtain a conservative estimate of the $T_{bc}$ discovery potential under a maximal-background assumption, we take the $1\sigma$ upper bound of the LO SPS 4FNS prediction, $\sigma^{\mathrm{LO~SPS~4FNS}}_{B^\mp D^\pm,\mathrm{max}} = 0.65~\upmu\mathrm{b}$, and combine it with the DPS contribution. This defines the \textit{conservative maximal-background scenario}.

\begin{figure}[htbp]
    \centering
    \includegraphics[width=0.8\linewidth]{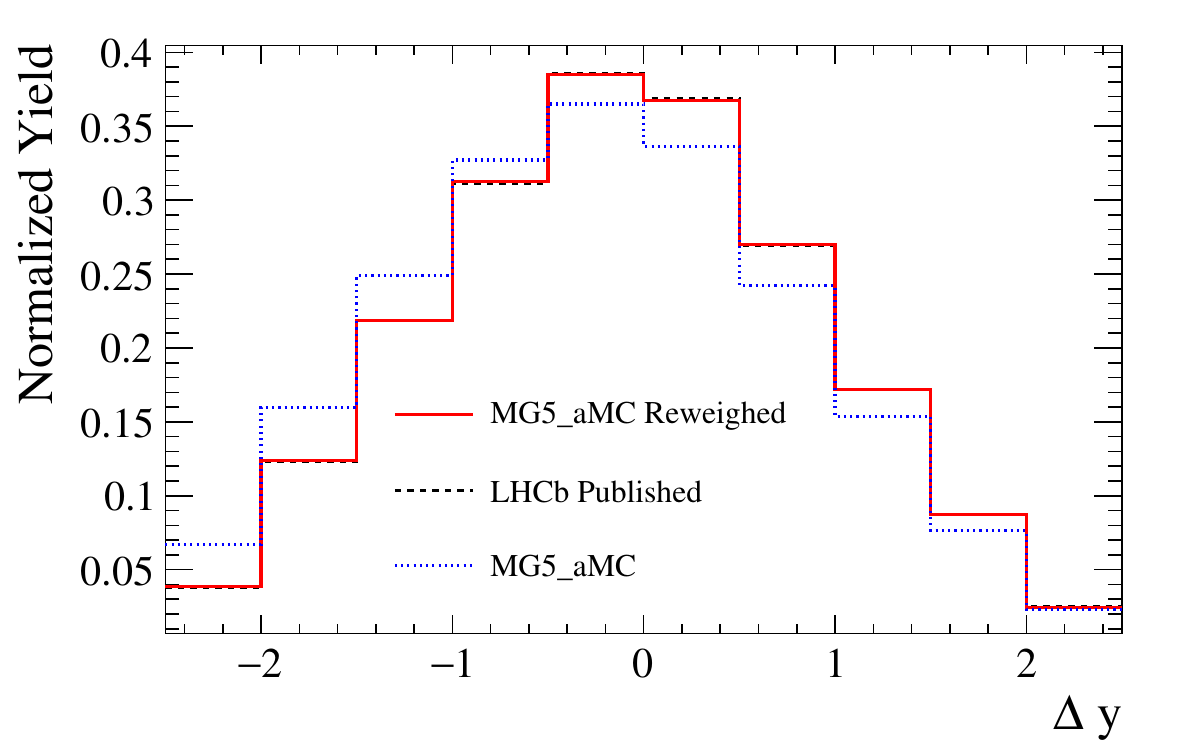}
    \caption{Comparison of the rapidity difference $\Delta y$ between the $B^\mp$ and $D^\pm$ mesons in the DPS background. The black dashed line represents the published LHCb Run~2 data, while the red solid and blue dotted lines correspond to the reweighted and unweighted DPS simulations, respectively. Results are shown for $\sigma_{\rm eff} = 15~\mathrm{mb}$.} 
    \label{fig:dps_fromLHCbPaper}
\end{figure}

To simulate the DPS background, two NLO event samples for the processes $pp \to b\bar{b}+X$ and $pp \to c\bar{c}+X$ are generated independently. The DPS $BD$ sample is obtained by randomly pairing events from the single-$B^\mp$ and single-$D^\pm$ samples. 
In the DPS approximation, the kinematic properties of the $B$ and $D$
mesons are largely governed by their respective single-particle
distributions. However, the single-particle rapidity spectra predicted
by $\mgshort$ may differ from those observed in data.
To reduce this model dependence, the DPS sample is reweighted
on an event-by-event basis according to the rapidity difference
$\Delta y = y_B - y_D$, using a reference $\Delta y$ distribution
constructed from the LHCb Run~2 measurements of the $B$- and $D$-meson
single-particle $d\sigma/dy$ spectra~\cite{LHCb:2017vec,
LHCb:2015swx}, as illustrated in Fig.~\ref{fig:dps_fromLHCbPaper}.

The total fiducial cross section of the DPS process is
\begin{equation}
\sigma_{B^\mp D^\pm}^{\mathrm{DPS}} = \frac{1}{2} \times \frac{\sigma_{B^\mp} \cdot \sigma_{D^\pm}}{\sigma_{\rm eff}}=\left(\frac{15~\mathrm{mb}}{\sigma_{\mathrm{eff}}}\right)\left(2.4\pm 0.3\right)~\mathrm{\upmu b}\,,
\end{equation}
where $\sigma_{B^\mp} = (86.6 \pm 6.4)~\upmu{\rm b}$~\cite{LHCb:2017vec} and $\sigma_{D^\pm} = (834 \pm 78)~\upmu{\rm b}$~\cite{LHCb:2015swx}, measured using the same fiducial cuts as in eq.~\eqref{eq:FDcuts}. The factor of $1/2$ accounts for the fact that only the charge-neutral combinations $B^-D^+$ and $B^+D^-$ are considered, since $\sigma_{B^\mp}$ and $\sigma_{D^\pm}$ include charge-conjugate contributions. The effective DPS cross section $\sigma_{\rm eff}$ is not precisely known but has been extracted from LHCb measurements of other heavy-flavor processes, such as $\Upsilon+D$~\cite{LHCb:2015wvu}, $J/\psi+\Upsilon$~\cite{LHCb:2023qgu}, and $J/\psi+D$~\cite{LHCb:2012aiv,Shao:2020kgj}, with typical values in the range $5$-$30$ mb. In this analysis, we consider representative values $\sigma_{\mathrm{eff}}=5,15$, and $30$ mb.

\begin{figure*}[htbp]
    \centering
    \includegraphics[width=0.3\linewidth]{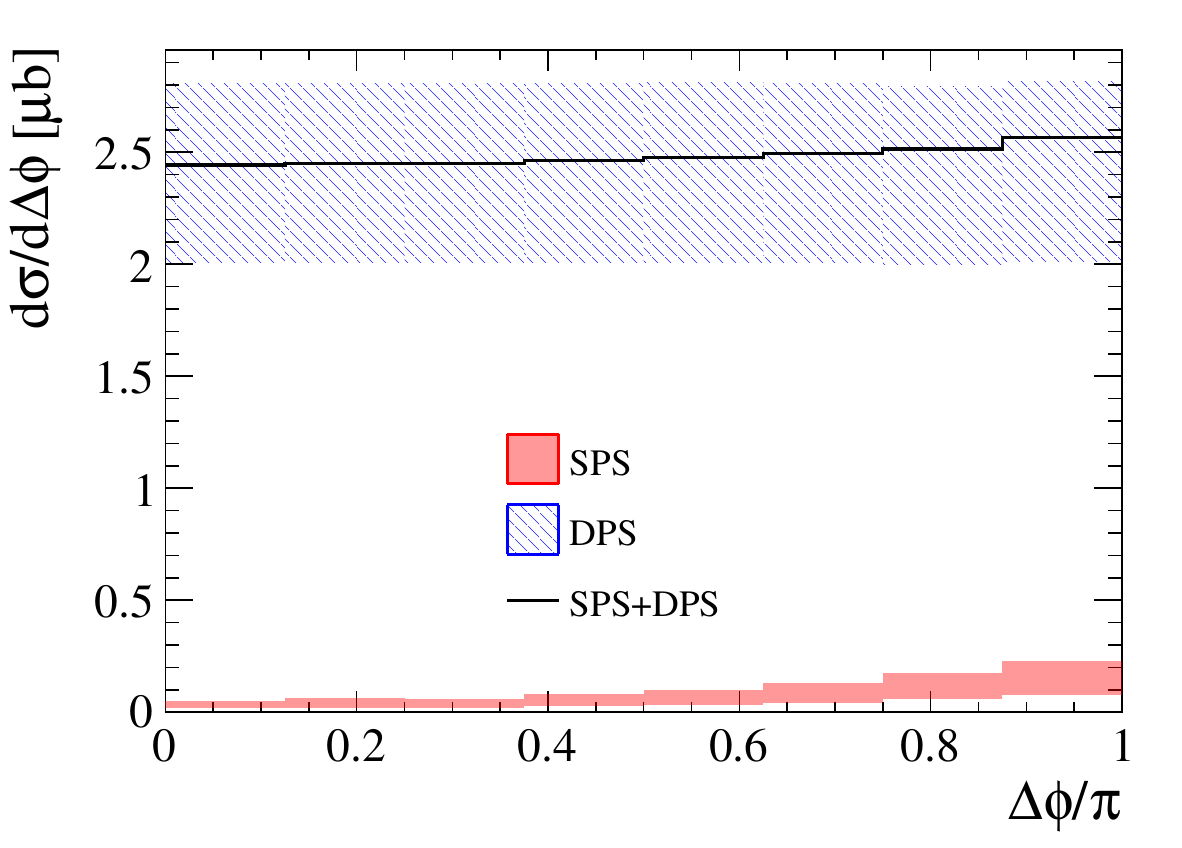}
    \includegraphics[width=0.3\linewidth]{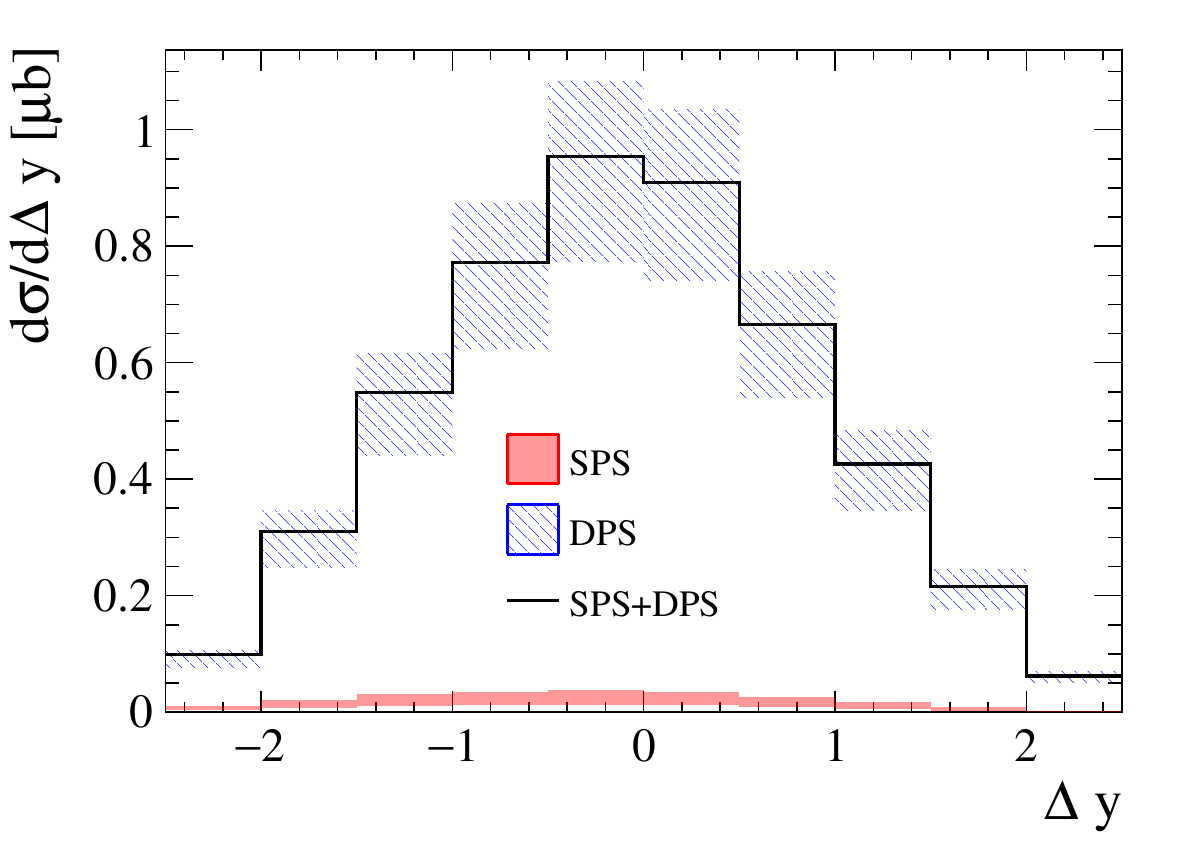}
    \includegraphics[width=0.3\linewidth]{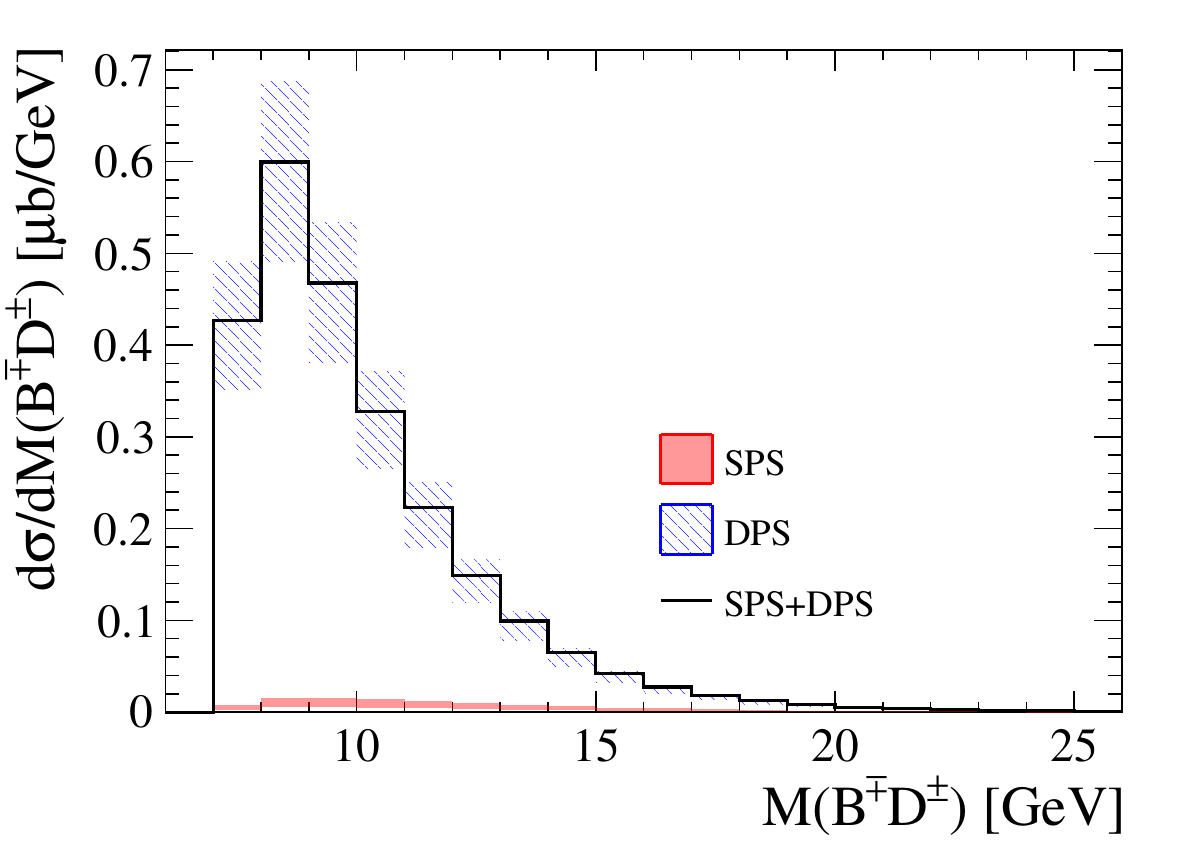}\\
    \includegraphics[width=0.3\linewidth]{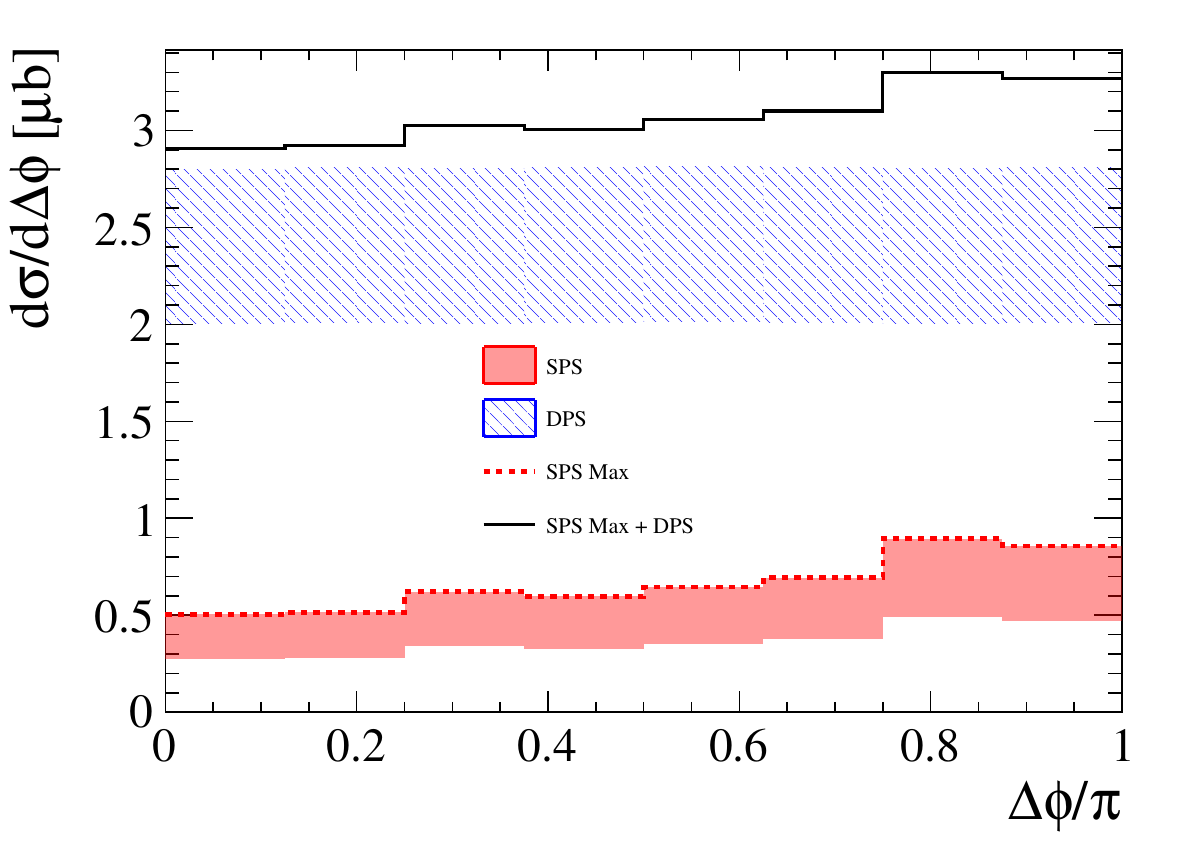}
    \includegraphics[width=0.3\linewidth]{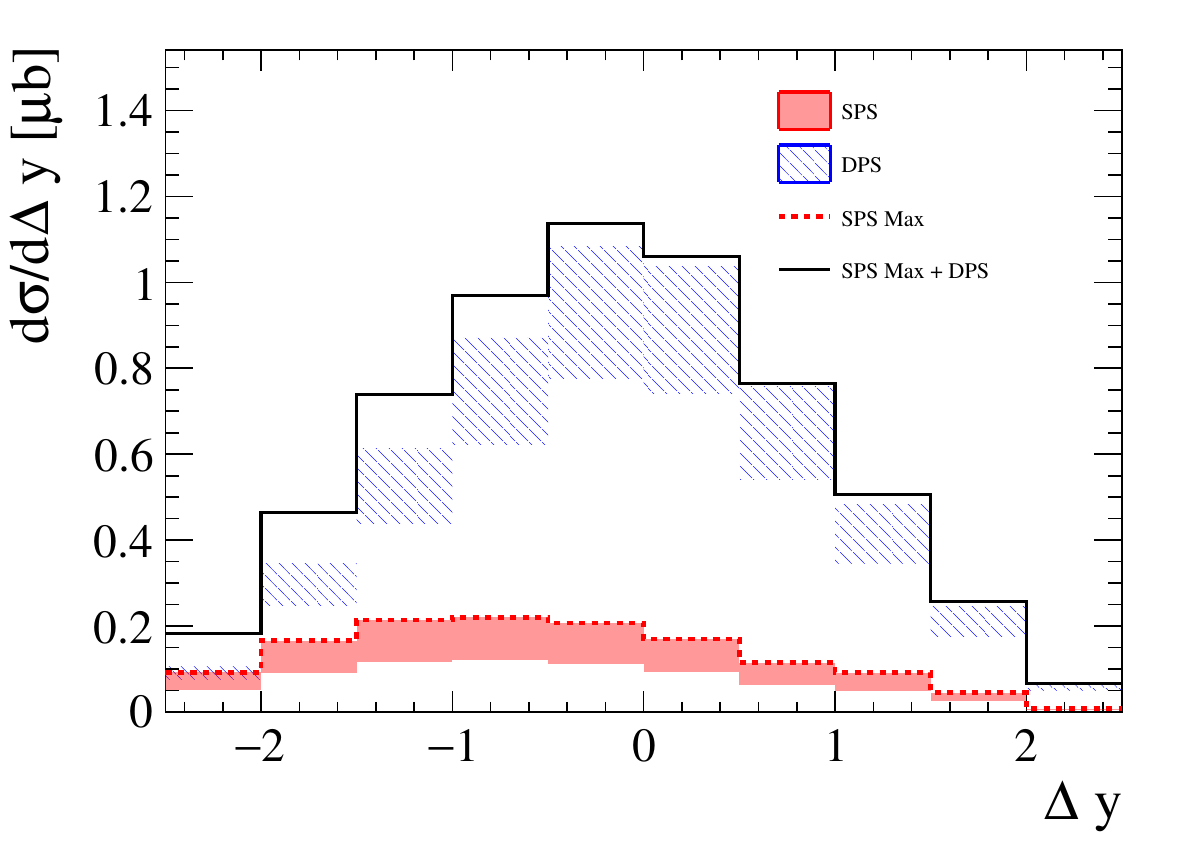}
    \includegraphics[width=0.3\linewidth]{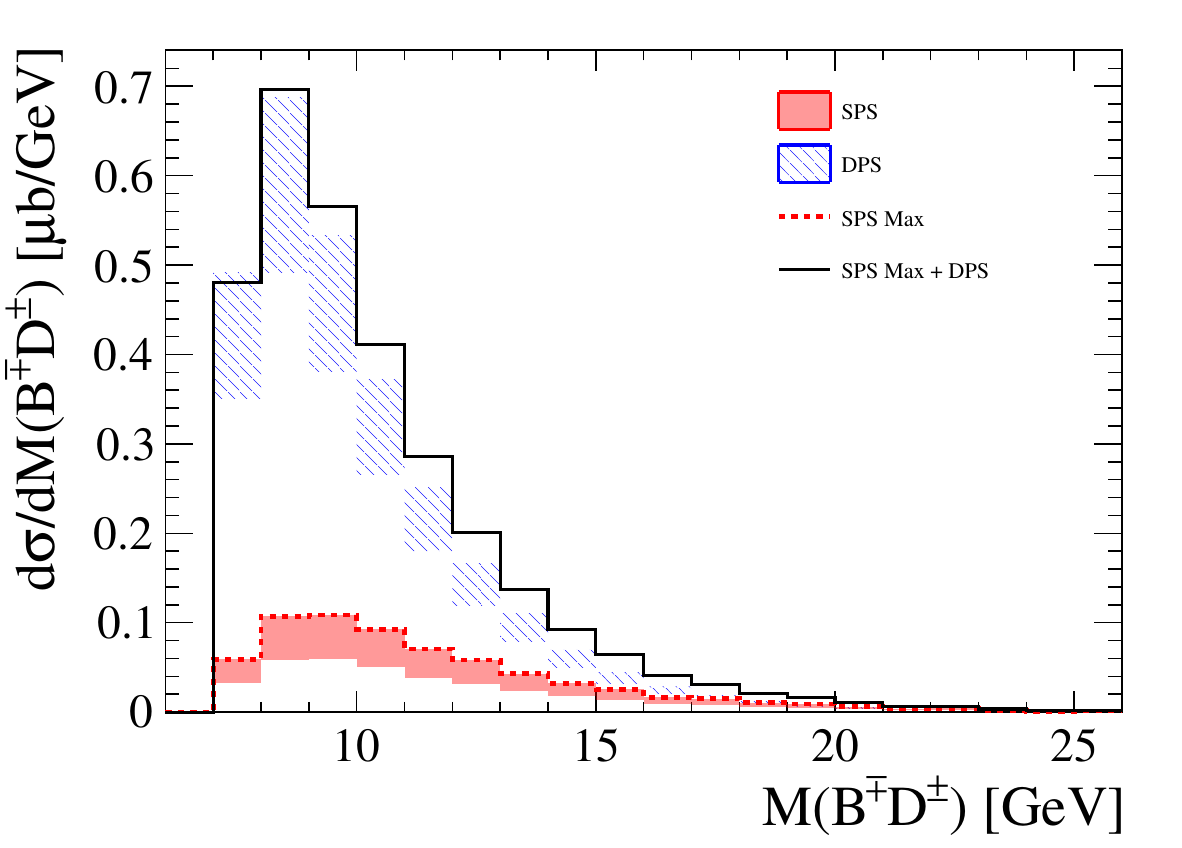}\\
    \caption{Differential cross sections for associated $B^\mp D^\pm$ production via SPS and DPS. Left: $\Delta\phi$ distribution; middle: $\Delta y$ distribution; right: invariant-mass $M(B^\mp D^\pm)$ distribution, shown for $\sigma_{\mathrm{eff}} = 15$~mb. The upper (lower) panels show the SPS prediction at NLO in the 3FNS (LO in the 4FNS). The SPS contribution is displayed as a red shaded band, with uncertainties dominated by renormalization- and factorization-scale variations, while the DPS contribution is displayed as a blue hatched band. In the upper panels, the black solid line denotes the combined SPS+DPS central prediction, which defines the baseline background scenario. In the lower panels, the red dashed line indicates the maximal SPS contribution, taken from the upper bound of the LO 4FNS prediction, while the black solid line shows the corresponding SPS (max)+DPS prediction, which defines the conservative maximal background scenario.}
    \label{fig:sps_dps_eff15}
\end{figure*}

Differential cross sections for associated $B^\mp D^\pm$ production via SPS, DPS, and their combination (SPS+DPS) are shown in Fig.~\ref{fig:sps_dps_eff15}. The upper panels correspond to the NLO SPS prediction in the 3FNS, while the lower panels show the LO SPS prediction in the 4FNS. The shaded bands denote the combined statistical and systematic uncertainties. For both SPS and DPS, statistical uncertainties are estimated from the event samples. The DPS systematic uncertainty originates from the uncertainties on $\sigma_{B^\mp}$ and $\sigma_{D^\pm}$, whereas the SPS systematic uncertainty is dominated by variations of the renormalization and factorization scales in the perturbative calculations. 

In the SPS mechanism, the azimuthal angle difference distribution, $\Delta\phi=|\phi(B)-\phi(D)|$, exhibits a prominent peak near $\pi$, indicating that $B^\mp$ and $D^\pm$ mesons are predominantly produced in a back-to-back configuration in the transverse plane. In contrast, the DPS mechanism yields a nearly uniform $\Delta \phi$ distribution, reflecting the absence of strong azimuthal correlations between the two mesons. Similarly, the rapidity gap $\Delta y$ shows distinct features for the two mechanisms. SPS events produce an asymmetric $\Delta y$ distribution, with the $B$ meson typically more central in rapidity than the $D$ meson, whereas DPS events result in a nearly symmetric distribution around $\Delta y = 0$, consistent with largely uncorrelated production of the two mesons. The invariant-mass distribution $M(B^\mp D^\pm)$ further highlights these differences. SPS events, dominated by back-to-back configurations, lead to a relatively broad distribution with a pronounced high-mass tail, while DPS events produce a narrower peak, though occasional high-momentum combinations can generate a modest tail. Overall, these observables illustrate the distinct kinematic patterns of SPS and DPS and provide several handles for disentangling their respective contributions.

Within the LHCb kinematic acceptance, associated $B^\mp D^\pm$ production is dominated by the DPS mechanism when using the nominal NLO SPS prediction in the 3FNS, under which the SPS contribution is negligible. Only when adopting the conservative upper-bound LO SPS calculation in the 4FNS does the SPS contribution become comparable to DPS in certain kinematic regions.

These differential distributions, based on the kinematic cuts defined in eq.~\eqref{eq:FDcuts}, will be used as the background for the $T_{bc}$ search in subsequent analyses.

\subsection{Generation of the $T_{bc}$ Signal}
\label{sec:Tbc_signal_generation}

In this analysis, the $T_{bc}$ signal is generated assuming specific values for its mass $m(T_{bc})$, width $\Gamma(T_{bc})$, and production cross section $\sigma(T_{bc})$, motivated by theoretical expectations. Signal samples are produced for different mass and width hypotheses. The $T_{bc}$ signal is modeled with a Breit–Wigner (BW) line shape to describe its natural width and is subsequently convolved with the detector response function to obtain the differential cross section.

The differential cross section for $T_{bc}$ production can be written as
\begin{equation}
\frac{\mathrm{d}\sigma_{T_{bc}}}{\mathrm{d}m} = \sigma(T_{bc}) \int_{-\infty}^{+\infty} 
\mathrm{BW}(m') \cdot G(m - m', \sigma_\mathrm{res}) \, \mathrm{d}m'\,,
\end{equation}
where $\sigma(T_{bc}) \equiv \sigma(pp \to T_{bc} + X)+\sigma(pp \to \overline{T}_{bc} + X)$ denotes the total production cross section for the $T_{bc}$ and $\overline{T}_{bc}$ states within the LHCb acceptance ($2<\eta(T_{bc})<4.5$)~\footnote{To obtain the inclusive cross section, or the inclusive limit on $\sigma(T_{bc}) \times \mathcal{B}$, over the full phase space (i.e., without fiducial cuts), the fiducial results should be divided by the LHCb acceptance. The acceptance is estimated to be $22–24\%$ based on simulations of $\Xi_{bc}$ and $B_c$ production and is assumed to be similar for $T_{bc}$ production.}, and $m\equiv M(B^\mp D^\pm)$ denotes the invariant mass of the reconstructed $B^{\mp}D^{\pm}$ pair. The nonrelativistic BW distribution is given by
\begin{equation}
\mathrm{BW}(m') =
\frac{1}{2\pi}
\frac{\Gamma(T_{bc})}{(m' - m(T_{bc}))^2 + (\Gamma(T_{bc})/2)^2}\,.
\end{equation}
The detector response function $G$ is modeled as a Gaussian distribution:
\begin{equation}
G(m-m',\sigma_\mathrm{res}) = \frac{1}{\sqrt{2\pi}\sigma_\mathrm{res}} \exp\Big[-\frac{(m-m')^2}{2\sigma_\mathrm{res}^2}\Big]\,,
\end{equation}
motivated by LHCb measurements of similar decay modes, such as $B_c^+ \to J/\psi D_s^+$ and $B_c^+ \to B_s^0 (\to D_s^- \pi^+)\pi^+$~\cite{LHCb:2020ayi}, for which the mass resolution is found to be in the range $4$-$7$ MeV. Given that the $T_{bc} \to B^- D^+$ decay chain, eq.~\eqref{eq:Tbcdecaychain}, involves multiple tracks and allows for intermediate mass constraints, we adopt a mass resolution of $\sigma_\mathrm{res} = 6~\mathrm{MeV}$ in the simulation.

In this study, we consider two representative mass hypotheses for the $T_{bc}$, $m(T_{bc}) = 7167~\mathrm{MeV}$ and $m(T_{bc}) = 7229~\mathrm{MeV}$, motivated by theoretical predictions for exotic hadron states with quantum numbers $J^P=0^+$~\cite{Cheng:2020wxa, Eichten:2017ffp}. These values correspond to possible $T_{bc}$ configurations and span a reasonable range consistent with current theoretical expectations for heavy tetraquarks. For these mass hypotheses, we consider several representative values of the width and production cross section. The width values $\Gamma(T_{bc}) = 0.5,~5~\mathrm{MeV}$ are used for $m(T_{bc}) = 7167~\mathrm{MeV}$, while $\Gamma(T_{bc}) = 10,~40~\mathrm{MeV}$ are adopted for $m(T_{bc}) = 7229~\mathrm{MeV}$. These choices are motivated by comparisons with similar exotic hadrons: the narrow width of the $T_{cc}^+$ ($\Gamma(T_{cc}^+) \sim 0.4~\mathrm{MeV}$)~\cite{LHCb:2021vvq}, and the broader widths of states such as $Z_c(3900)$ and $Z_b(10610)$, which have widths of order $10$-$30$ MeV~\cite{BESIII:2013ris, Belle:2015upu}.

The theoretical prediction for the $T_{bc}$ production cross section is highly uncertain, with different approaches yielding results that can differ by two orders of magnitude, ranging from a very conservative value of $0.3$~nb~\cite{Chen:2011jtl} to an optimistic estimate of $103$~nb~\cite{Ali:2018xfq}. In addition to the two spanning a wide range, we also consider an intermediate estimate using
\begin{equation}
\sigma(T_{bc}) \approx \sigma(T_{cc}^+) \times \frac{\sigma(\Xi_{bc})}{\sigma(\Xi_{cc})} \approx 45 \times 0.4 \approx 18~\mathrm{nb}\,,
\end{equation}
where $\sigma(T_{cc}^+) = 45 \pm 20~\mathrm{nb}$ is the prompt production cross section of the $T_{cc}^+$ tetraquark at LHCb within the typical acceptance ($2 < p_T/\mathrm{GeV} < 20$, $2.0 < y < 4.5$)~\cite{Ali:2024hzp}, and $\sigma(\Xi_{bc})/\sigma(\Xi_{cc}) \approx 0.4$ is adopted from theoretical calculations of doubly heavy baryon production at LHC energies~\cite{Zhang:2011hi}. In summary, all three representative values (optimistic: $103$~nb, intermediate: $18$~nb, conservative: $0.3$~nb) are considered in this study.

Based on these assumptions, we simulate the $T_{bc}$ signal and evaluate the discovery potential across different parameter combinations. Since the $T_{bc}$ resonance is narrow and the background is largely smooth in the $B^\mp D^\pm$ invariant mass spectrum, signal-background interference is expected to be small. In particular, the interference effect decreases as the signal width $\Gamma(T_{bc})$ decreases and can be safely neglected for the narrowest width considered ($\Gamma(T_{bc}) = 0.5~\mathrm{MeV}$). Therefore, we neglect signal-background interference in our study.

\subsection{Signal and Background Yields}

In this analysis, once the cross-section distributions for the signal and background have been constructed, the numbers of signal and background events are estimated based on the decay chain in eq.~\eqref{eq:Tbcdecaychain}. The event yields for both signal and background are then computed using the standard event-counting formula:
\begin{equation}
    N = \sigma \times \mathcal{L}_{\mathrm{int}} \times \mathcal{B} \times \varepsilon\,,\label{eq:Nevent} 
\end{equation}
where
\begin{itemize}
  \item $\sigma$ denotes the production cross section. For the signal, $\sigma=\sigma(T_{bc})$, whereas for the background it is the sum of the SPS and DPS contributions: $\sigma = \sigma_{B^\mp D^\pm}^{\mathrm{DPS}} + \sigma_{B^\mp D^\pm}^{\mathrm{SPS}}$. In this study, we neglect the dependence of $\sigma$ on the $pp$ center-of-mass energy between 13~TeV and 14~TeV.
  \item $\mathcal{L}_{\mathrm{int}}$ is the integrated luminosity. LHCb has already collected approximately $6~\mathrm{fb}^{-1}$ of $pp$ collision data in Run~2 at $\sqrt{s}=13$ TeV. The Run~1 dataset is not included in this study, as its relatively small integrated luminosity and lower center-of-mass energies (7 and 8 TeV) make a negligible contribution to the overall sensitivity. By the end of Run~4, the total integrated luminosity from Runs~2, 3, and 4 is expected to reach $\sim 50~\mathrm{fb}^{-1}$, and by the end of Run~5 it is expected to reach $\sim 300~\mathrm{fb}^{-1}$. Therefore, in this study we scan $\mathcal{L}_{\mathrm{int}}$ over the range $5$-$300~\mathrm{fb}^{-1}$.
  \item $\mathcal{B}$ denotes the branching fraction. The relevant values are $\mathcal{B}(B^\pm \to J/\psi K^\pm) = (1.020 \pm 0.019) \times 10^{-3}$, $\mathcal{B}(D^\pm \to K^\mp \pi^\pm \pi^\pm) = (9.38 \pm 0.16) \times 10^{-2}$, and $\mathcal{B}(J/\psi \to \mu^+ \mu^-) = (5.961 \pm 0.033) \times 10^{-2}$~\cite{ParticleDataGroup:2024cfk}. For the signal decay $T_{bc} \to B^- D^+$, we adopt a default branching fraction $\mathcal{B}(T_{bc} \to B^- D^+) = 0.5$. This choice is motivated by the expectation that the $J^P=0^+$ $T_{bc}$ state decays predominantly into the two open-heavy-flavor channels, $\bar{B}^0 D^0$ and $B^-D^+$, with comparable rates. In the absence of detailed theoretical calculations, this symmetric assignment serves as a reasonable reference point, and the associated uncertainty is effectively accounted for by the scan over $\sigma(T_{bc}) \times \mathcal{B}$ presented in Tables~\ref{tab:SigmaBr_50invfb} and \ref{tab:SigmaBr_300invfb}.
  \item $\varepsilon$ denotes the overall event reconstruction and selection efficiency, including contributions from the detector geometric acceptance, track reconstruction, trigger, particle identification, and offline selection. For the decay $T_{bc} \to B^- D^+$, the efficiency is given by the product $\varepsilon_{B}\times \varepsilon_{D}$, with the $D$-meson trigger efficiency excluded, since the $BD$ candidates are triggered by the $B$ meson. The total efficiency is estimated using measurements of related decays~\cite{LHCb:2017vec, LHCb:2015swx}, yielding $\varepsilon_{BD} = (0.6 \pm 0.3)\%$; the $50\%$ relative uncertainty is a conservative estimate that accounts for potential correlations between the two efficiency factors and for variations across the full kinematic range and different data-taking conditions.
  
\end{itemize}

\begin{figure}[htbp]
    \centering
     \includegraphics[width=0.9\linewidth] {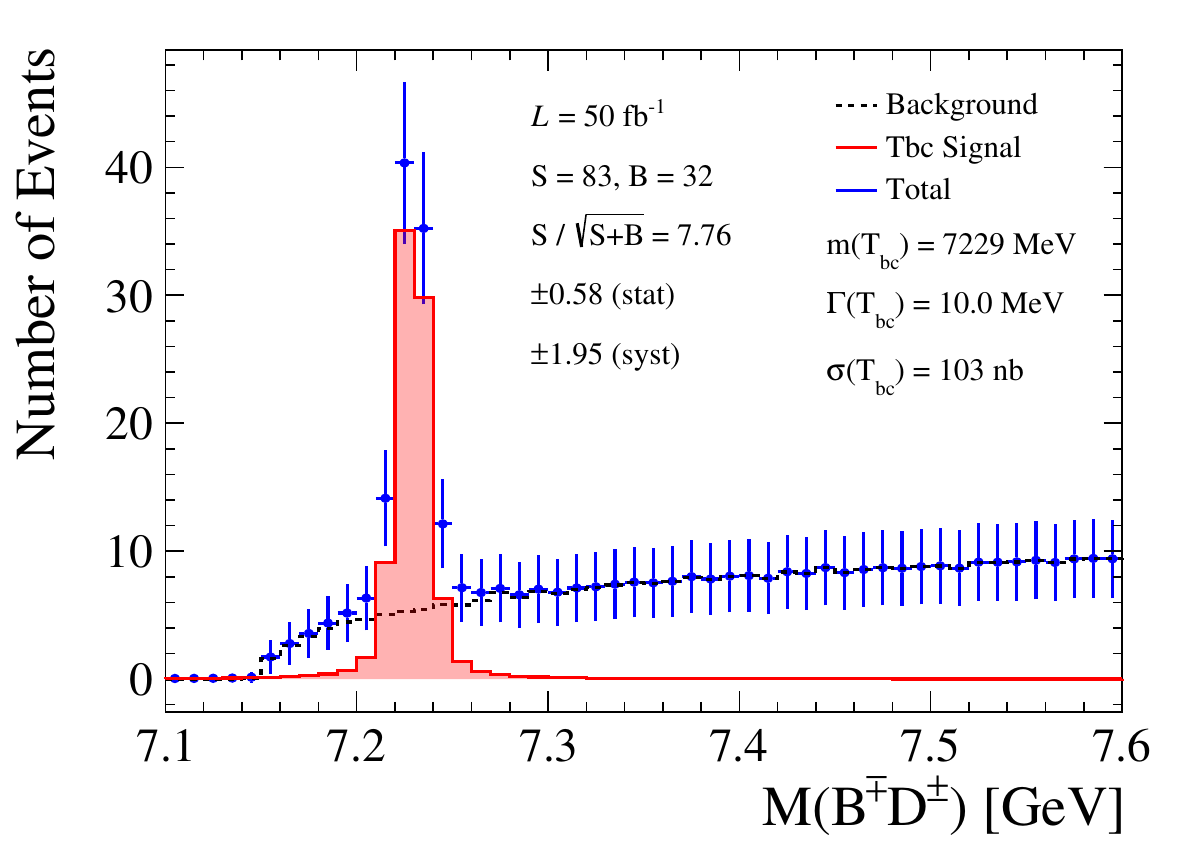}
    \caption{Combined distribution of the $T_{bc}$ signal and background in the $BD$ invariant-mass spectrum, assuming an effective DPS cross section $\sigma_{\mathrm{eff}} = 15~\mathrm{mb}$, $m(T_{bc}) = 7229~\mathrm{MeV}$, and $\Gamma(T_{bc}) = 10~\mathrm{MeV}$. The black curve represents the $B^\mp D^\pm$ background, the red curve the $T_{bc}$ signal, and the blue curve their sum.}
    \label{fig:Tbc_Yield}
\end{figure}

Figure~\ref{fig:Tbc_Yield} shows an example of the invariant-mass spectrum of the reconstructed $B^\mp D^\pm$ pairs at an integrated luminosity of $50~\mathrm{fb}^{-1}$, assuming a $T_{bc}$ production cross section $\sigma(T_{bc}) = 103~\mathrm{nb}$ and an effective DPS cross section $\sigma_{\mathrm{eff}} = 15~\mathrm{mb}$. In this case, the $T_{bc}$ signal appears as a distinct resonance peak at the expected mass.

\subsection{Discovery Potential}

After obtaining the signal and background yields, $N_{\mathrm{Sig}}$ and $N_{\mathrm{Bkg}}$, under different parameter assumptions, the significance of the $T_{bc}$ signal is estimated using the standard formula:
\begin{equation}
    Z = \frac{N_{\mathrm{Sig}}}{\sqrt{N_{\mathrm{Sig}} + N_{\mathrm{Bkg}}}}\,,\label{eq:Zsig}
\end{equation}
where $N_{\mathrm{Sig}}$ and $N_{\mathrm{Bkg}}$ denote the numbers of signal and background events within a $\pm 3\sigma_{\mathrm{eff~res}}$ mass window, as evaluated using eq.~\eqref{eq:Nevent}. The effective mass resolution is defined as 
\begin{equation}
\sigma_{\mathrm{eff~res}} = \sqrt{\left(\frac{\Gamma(T_{bc})}{2.35}\right)^2 + \sigma_{\mathrm{res}}^2}\,.
\end{equation}
The mass window is centered on the signal peak in the $B^\mp D^\pm$ invariant-mass spectrum.

In this study, we evaluate the discovery potential of $T_{bc}$ by scanning the significance $Z$ as a function of the integrated luminosity $\mathcal{L}_{\mathrm{int}}$ for different $T_{bc}$ mass and width hypotheses, as well as different values of the effective DPS cross section $\sigma_{\mathrm{eff}}$ that contribute to the background. This scan allows us to determine the minimum integrated luminosity required for a $5\sigma$ discovery of $T_{bc}$ under these assumptions.

In addition, we scan over $\sigma(T_{bc}) \times \mathcal{B}(T_{bc} \to B^- D^+)$ to evaluate how different assumed signal strengths affect the discovery significance. By calculating $Z$ for various integrated luminosities, we determine the minimum observable value of $\sigma(T_{bc}) \times \mathcal{B}(T_{bc} \to B^- D^+)$ corresponding to a $5\sigma$ threshold, providing a quantitative assessment of the $T_{bc}$ discovery potential under different experimental conditions.

\subsection{Systematic Uncertainties}
\label{sec:systematics}

In this study, several sources of systematic uncertainty affect the estimate of the significance $Z$. The branching fractions $\mathcal{B}(B^\pm \to J/\psi K^\pm)$, $\mathcal{B}(J/\psi \to \mu^+\mu^-)$, and $\mathcal{B}(D^+ \to K^-\pi^+\pi^+)$ are taken from ref.~\cite{ParticleDataGroup:2024cfk}, with relative uncertainties of $1.9\%$, $0.6\%$, and $1.7\%$, respectively. The detector efficiency, $\varepsilon = (0.6 \pm 0.3)\%$, corresponds to a $50\%$ relative uncertainty. These uncertainties are fully correlated between signal and background because both yields depend on the same inputs.

An additional uncertainty affects only the background yield and arises from uncertainties in the measured $B^\mp$ and $D^\pm$ production cross sections. Propagating the uncertainties on $\sigma_{B^\mp}$ and $\sigma_{D^\pm}$ yields an approximate $10\%$ relative uncertainty in the DPS background cross section. The normalization of the DPS cross section is also subject to significant uncertainty due to $\sigma_{\mathrm{eff}}$, which is included in our evaluation. The SPS background uncertainty is evaluated using the baseline and conservative scenarios described in Section~\ref{sec:bkg}.

The total systematic uncertainty on the significance $Z$ is computed by propagating these uncertainties, accounting for the correlation structure outlined above. Among the considered sources, the detector-efficiency uncertainty dominates. This procedure is applied to each parameter scenario and is reflected in all results presented in Section~\ref{sec:results}.

\section{RESULTS}
\label{sec:results}

\subsection{Minimum Integrated Luminosity for a $5\sigma$ Discovery at LHCb}

\begin{figure*}[htbp]
    \centering
    \includegraphics[width=0.335\linewidth]{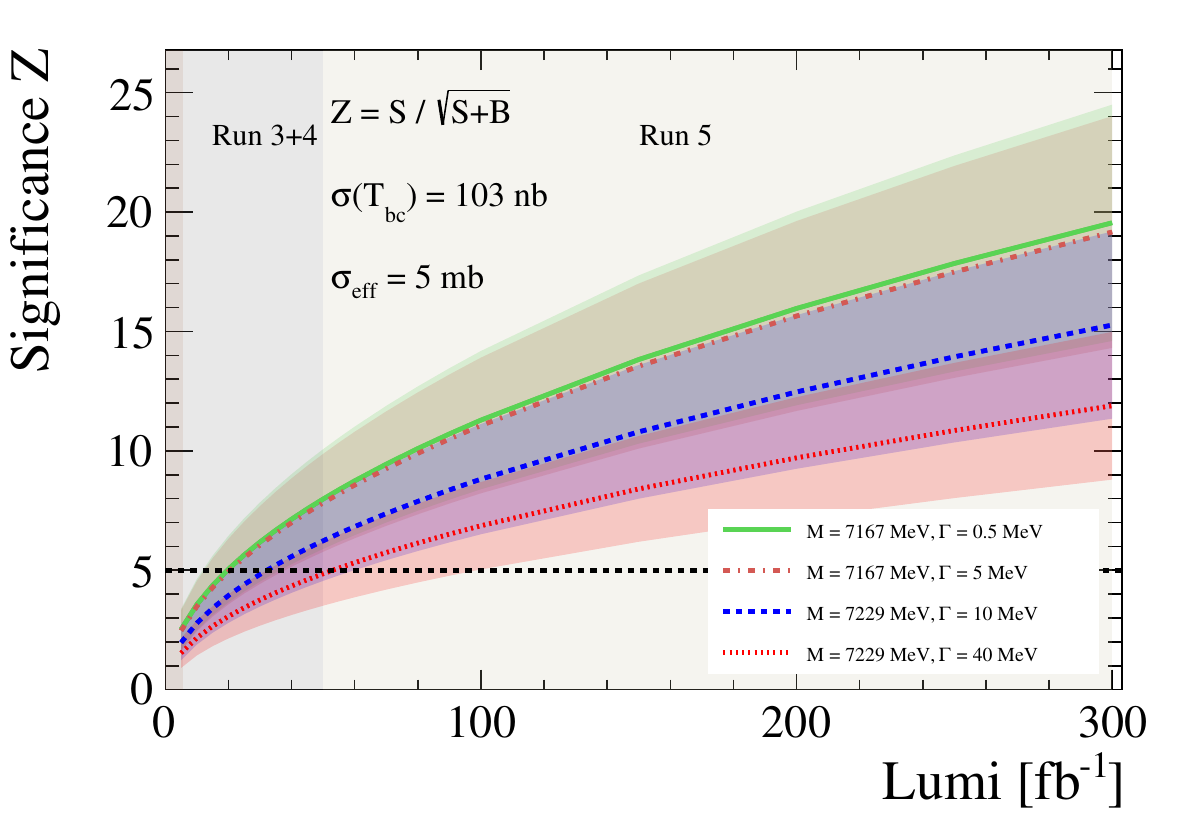}\hspace{-0.3cm}
    \includegraphics[width=0.335\linewidth]{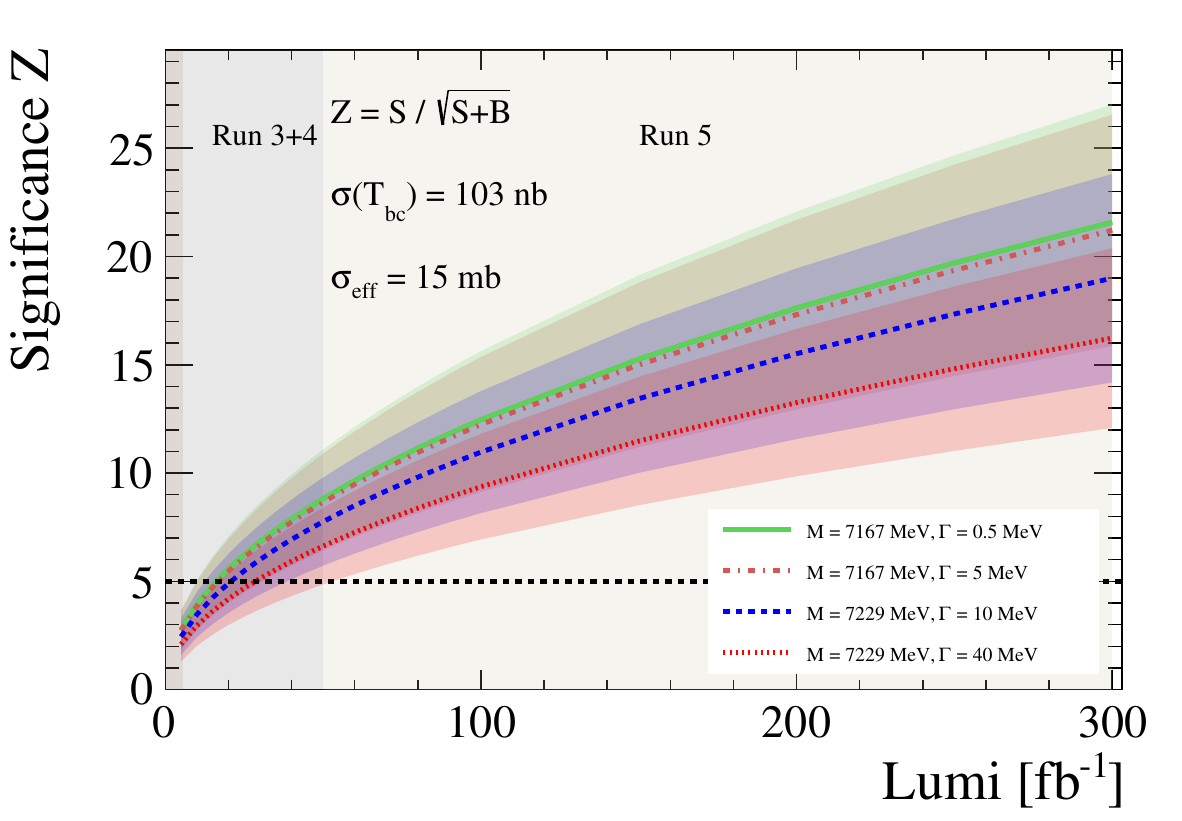}\hspace{-0.3cm}
    \includegraphics[width=0.335\linewidth]{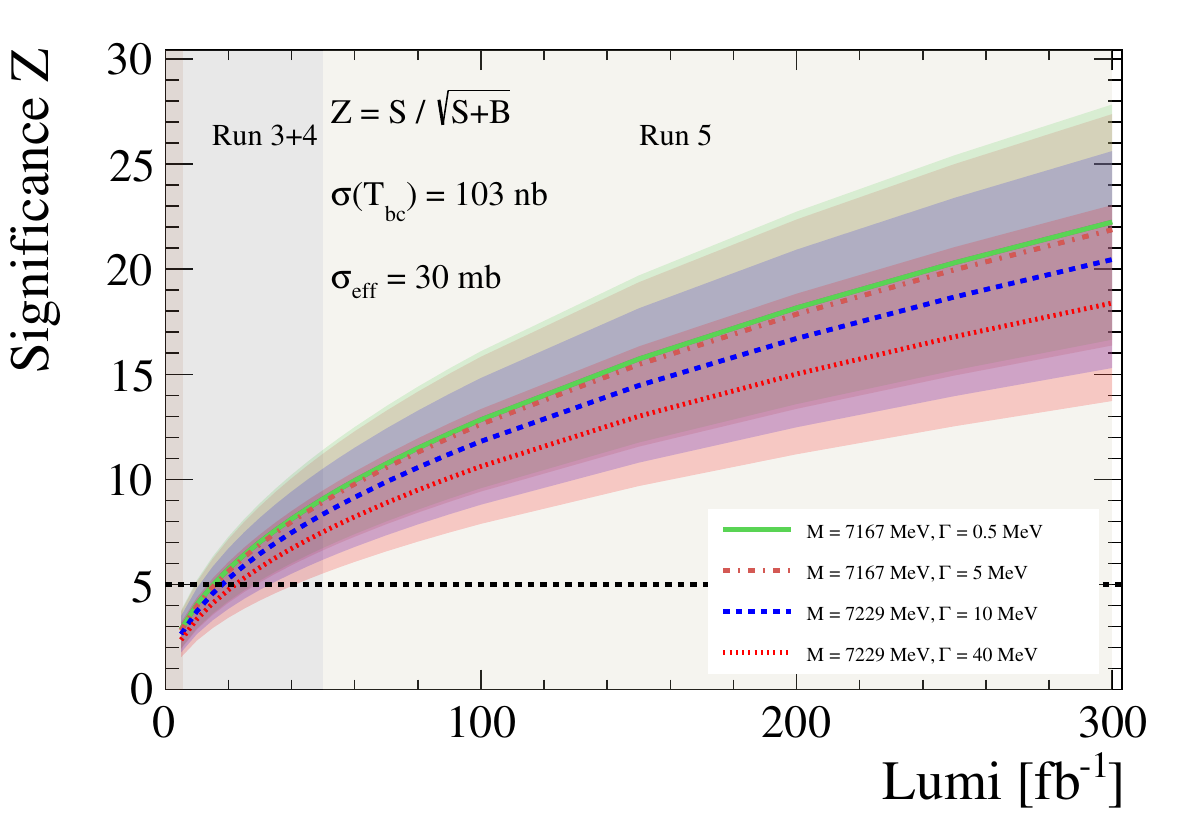}\\
    \includegraphics[width=0.335\linewidth]
    {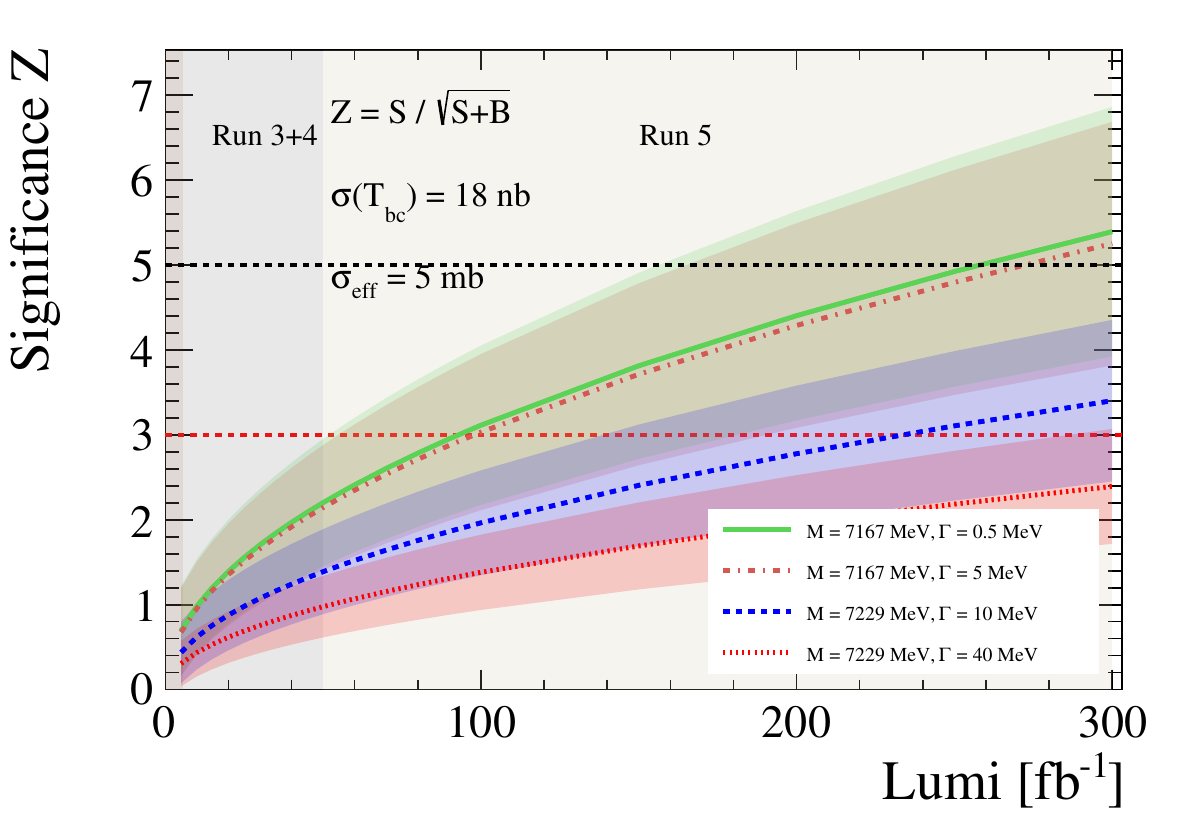}\hspace{-0.3cm}
    \includegraphics[width=0.335\linewidth]{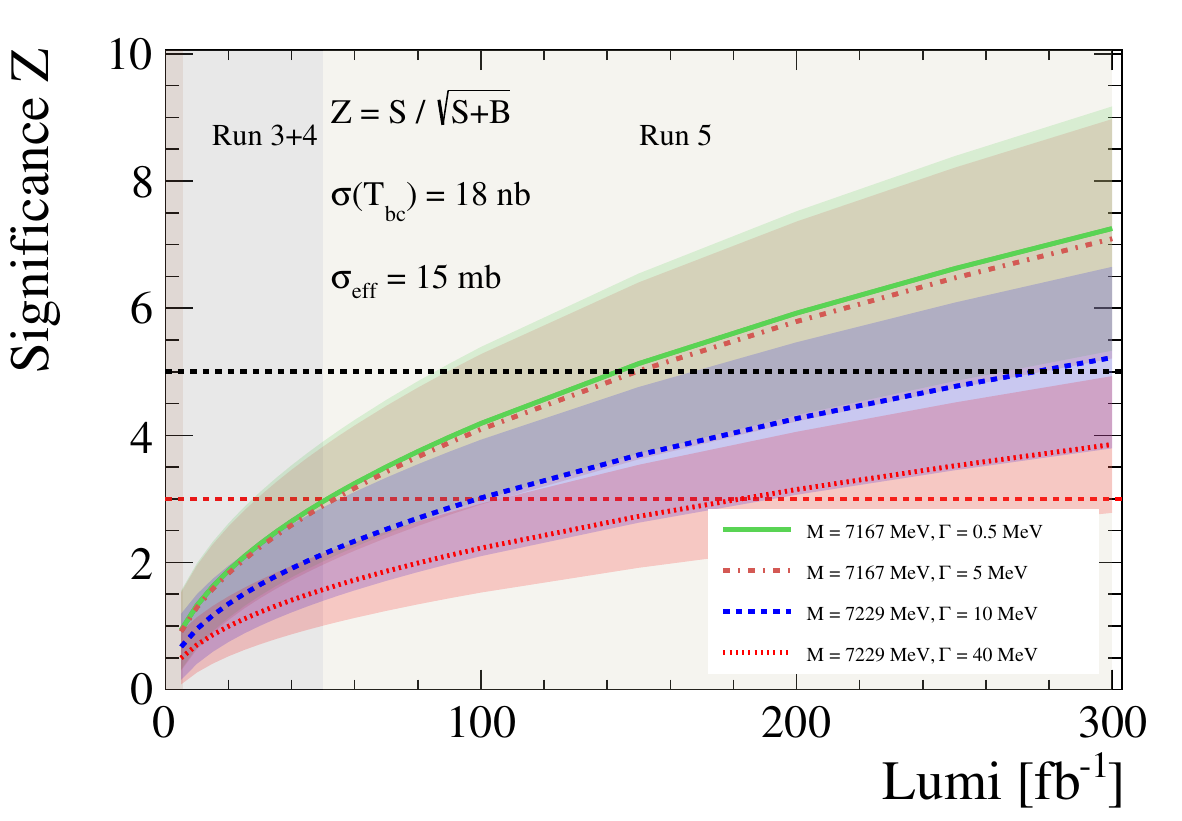}\hspace{-0.3cm}
    \includegraphics[width=0.335\linewidth]{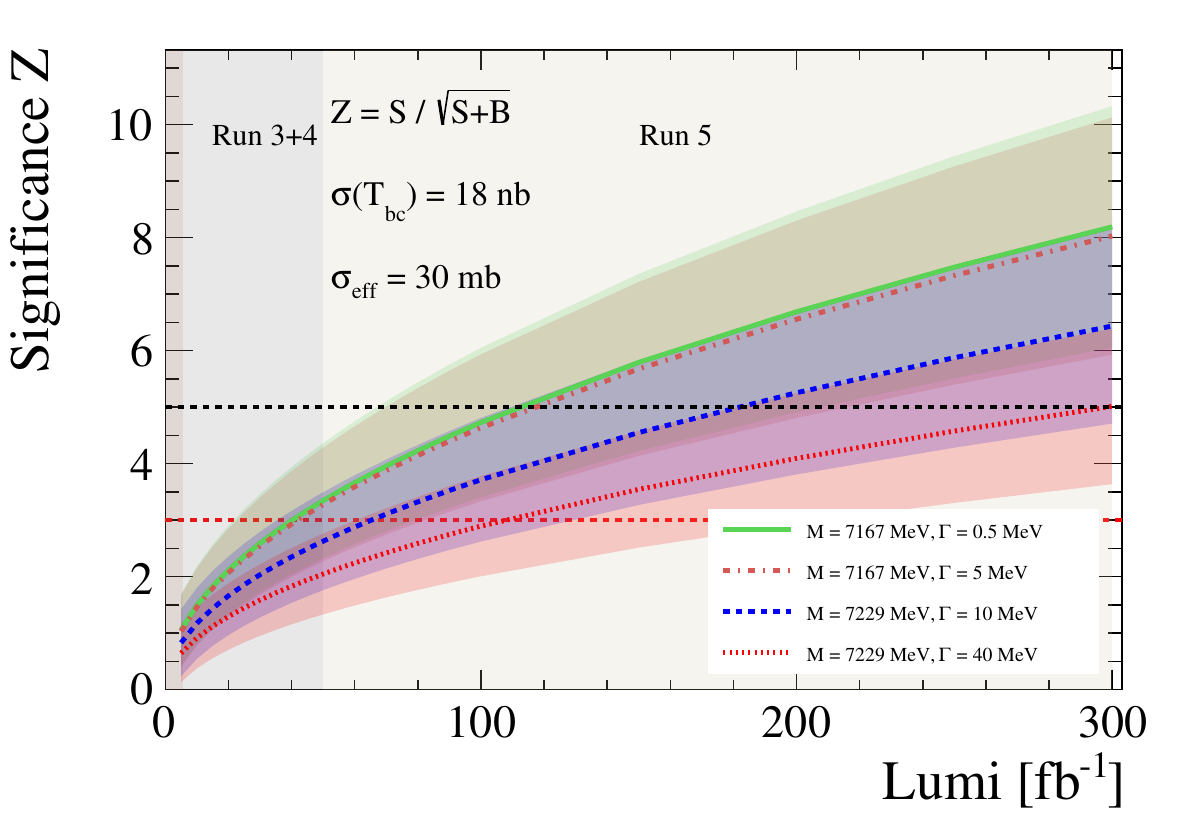}\\
    \includegraphics[width=0.335\linewidth]
    {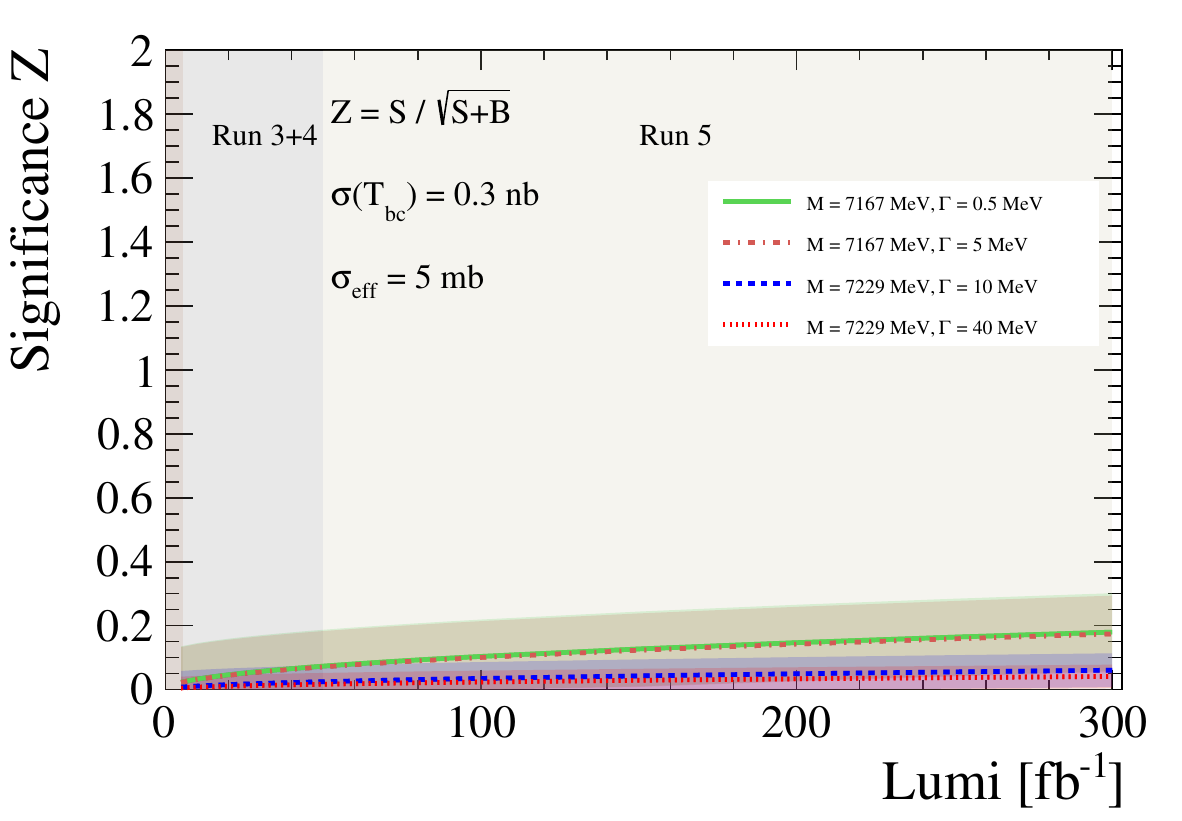}\hspace{-0.3cm}
    \includegraphics[width=0.335\linewidth]{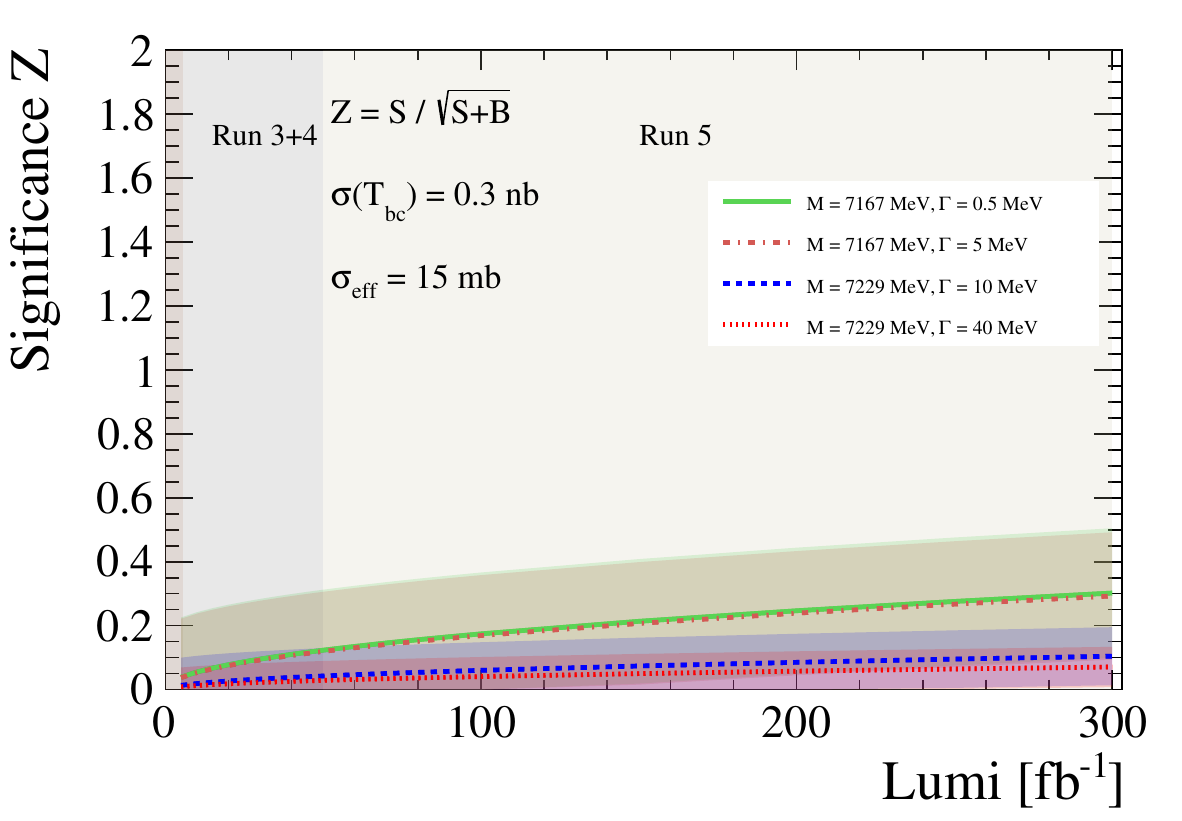}\hspace{-0.3cm}
    \includegraphics[width=0.335\linewidth]{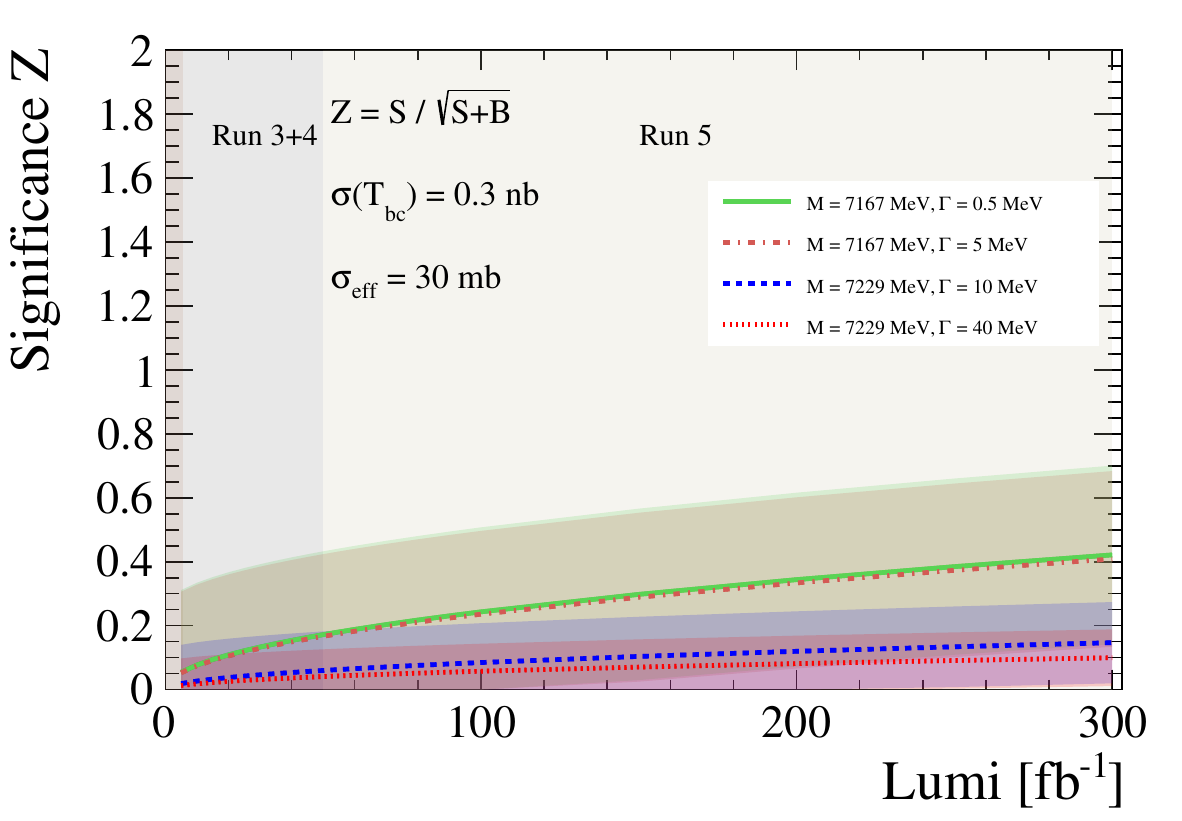}\\
    \caption{Shown is the statistical significance $Z$ of the $T_{bc}$ signal in the $BD$ invariant-mass spectrum as a function of the integrated luminosity $\mathcal{L}_{\mathrm{int}}$, for different $T_{bc}$ mass and width assumptions and for various values of the effective DPS cross section $\sigma_{\mathrm{eff}}$. The top, middle, and bottom panels correspond to production cross sections $\sigma(T_{bc}) = 103$ nb, $18$ nb, and $0.3~\mathrm{nb}$, respectively. A baseline background scenario (DPS + NLO SPS in the 3FNS) and a branching fraction $\mathcal{B}(T_{bc}\to B^-D^+)=0.5$ are assumed. The left, middle, and right panels correspond to $\sigma_{\mathrm{eff}}=5~\mathrm{mb}$, $15~\mathrm{mb}$, and $30~\mathrm{mb}$, respectively. Different colors denote different $T_{bc}$ mass and width assumptions: $M=7167~\mathrm{MeV}$ with $\Gamma=0.5,~5~\mathrm{MeV}$, and $M=7229~\mathrm{MeV}$ with $\Gamma=10,~40~\mathrm{MeV}$, where $M \equiv m(T_{bc})$ and $\Gamma \equiv \Gamma(T_{bc})$. The dashed line indicates the $5\sigma$ discovery threshold.}
    \label{fig:Tbc_discovery_potential}
\end{figure*}

Figure~\ref{fig:Tbc_discovery_potential} shows the statistical significance $Z$ of the $T_{bc}$ signal in the $BD$ invariant-mass spectrum as a function of integrated luminosity $\mathcal{L}_{\mathrm{int}}$, for different values of the effective DPS cross section $\sigma_{\mathrm{eff}}$ and for various assumptions on the $T_{bc}$ mass and width. The top, middle, and bottom panels correspond to production cross sections of $\sigma(T_{bc}) = 103$ nb, $18$ nb, and $0.3$ nb, respectively. We assume a baseline background scenario and take the branching fraction to be $\mathcal{B}(T_{bc}\to B^-D^+)=0.5$. The values of $\sigma_{\mathrm{eff}}$ are set to $5$, $15$, and $30$ mb, from left to right. As $\sigma_{\mathrm{eff}}$ increases, the DPS background cross section decreases, thereby enhancing the discovery potential for $T_{bc}$.

\begin{figure}[htbp]
    \centering
    \includegraphics[width=0.9\linewidth]{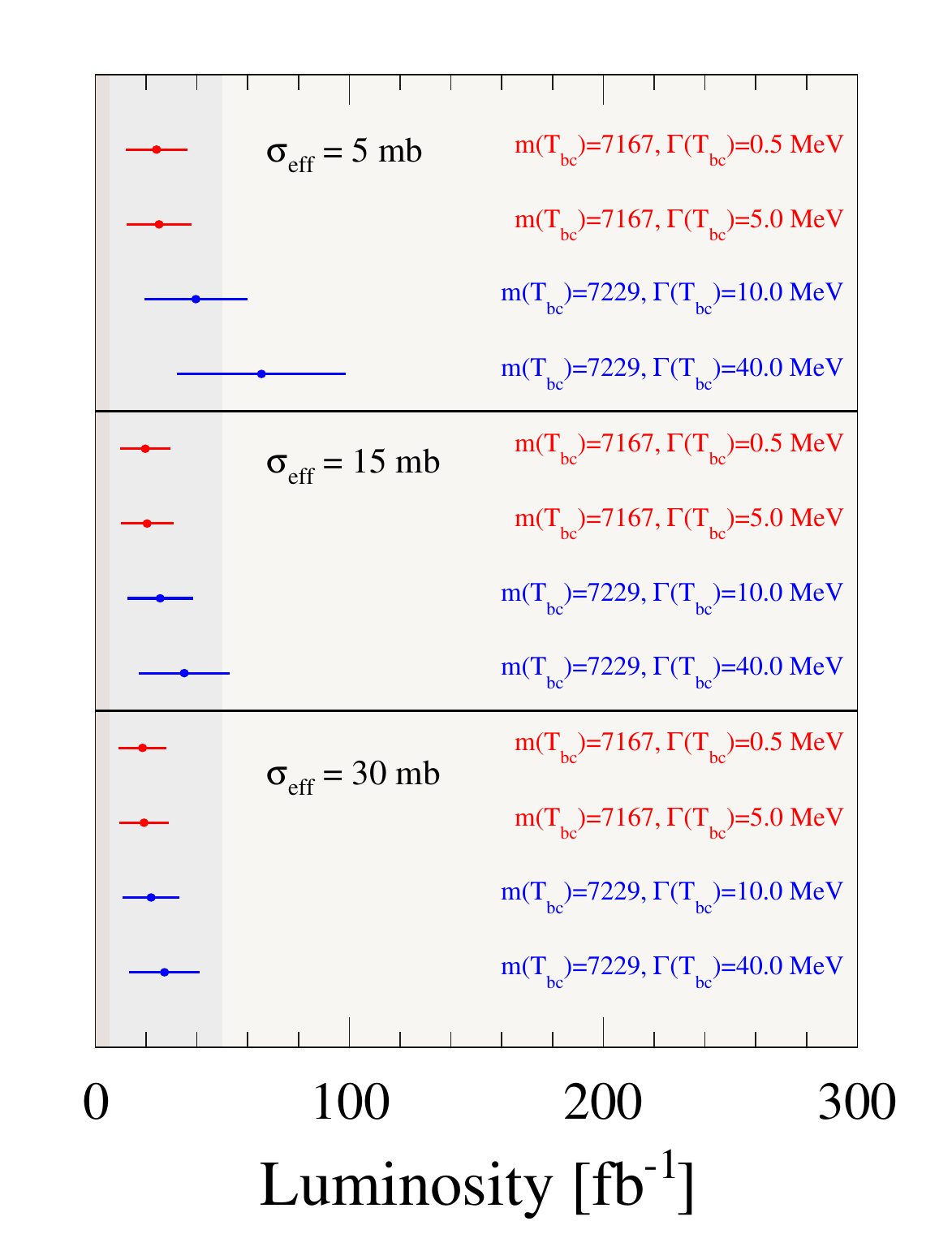}
    \caption{Discovery significance for $T_{bc}$ as a function of integrated luminosity for different $T_{bc}$ mass and width parameters, assuming the baseline background scenario with $\sigma(T_{bc}) = 103~\mathrm{nb}$ and $\mathcal{B}(T_{bc}\to B^-D^+)=0.5$.}
    \label{fig:lumifor5sigma}
\end{figure}

For the optimistic estimate of $\sigma(T_{bc})=103~\mathrm{nb}$ (top panels), Fig.~\ref{fig:lumifor5sigma} and Table~\ref{tab:lumifor5sigma} show the minimum integrated luminosity required for a $5\sigma$ discovery. For most parameter sets, this is achievable with $50~\mathrm{fb}^{-1}$ (Run~2 + Run~3 + Run~4). A comparison between baseline and conservative background scenarios indicates that this requirement is robust against SPS modeling uncertainties. For the intermediate estimate of $\sigma(T_{bc})=18~\mathrm{nb}$, the discovery potential is reduced relative to the optimistic case. Table~\ref{tab:lumifor5sigma_18nb} shows the minimum integrated luminosity required for $3\sigma$ evidence and $5\sigma$ discovery. At $300~\mathrm{fb}^{-1}$, corresponding to the full LHCb dataset up to Run~5, the significance exceeds $3\sigma$ for most parameter sets, and only the most favorable scenarios reach $5\sigma$. In the most pessimistic scenario with $\sigma(T_{bc}) = 0.3~\mathrm{nb}$, the statistical significance remains well below $3\sigma$ for all parameter sets considered, even with $300~\mathrm{fb}^{-1}$. Under this pessimistic assumption, no significant signal is expected.

\begin{table*}[b]
\caption{The integrated luminosity required for a $5\sigma$ discovery is evaluated under different $T_{bc}$ parameter assumptions, assuming $\sigma(T_{bc})=103~\mathrm{nb}$ and $\mathcal{B}(T_{bc}\to B^{-} D^{+})=0.5$. Both the baseline and a conservative, maximal-background scenario are considered. The quoted uncertainties include both statistical and systematic contributions.}
\label{tab:lumifor5sigma}
\begin{ruledtabular}
\begin{tabular}{cccc}
DPS $\sigma_{\mathrm{eff}}$ [mb] & m($T_{bc}$), $\Gamma$($T_{bc}$) [MeV] 
& $\mathcal{L}_{\mathrm{int}}$ [$\mathrm{fb}^{-1}$] (NLO SPS 3FNS) 
& $\mathcal{L}_{\mathrm{int}}$ [$\mathrm{fb}^{-1}$] (LO SPS 4FNS, max) \\
\hline
\multirow{4}{*}{5}      & 7167, 0.5 &  24 $\pm$ 12  &  25 $\pm$ 13\\
       & 7167, 5.0 &  25 $\pm$ 13  &  26 $\pm$ 13\\
       & 7229, 10  &  40 $\pm$ 20  &  40 $\pm$ 21\\
       & 7229, 40  &  66 $\pm$ 33  &  68 $\pm$ 34\\
\hline
\multirow{4}{*}{15}    & 7167, 0.5 &  20 $\pm$ 10  &  20 $\pm$ 10\\
       & 7167, 5.0 &  21 $\pm$ 10  &  21 $\pm$ 10\\
       & 7229, 10  &  26 $\pm$ 13  &  26 $\pm$ 13\\
       & 7229, 40  &  35 $\pm$ 18  &  37 $\pm$ 19\\
\hline
\multirow{4}{*}{30}     & 7167, 0.5 &  19 $\pm$ 9   &  19 $\pm$ 10\\
       & 7167, 5.0 &  19 $\pm$ 10  &  20 $\pm$ 10\\
       & 7229, 10  &  22 $\pm$ 11  &  23 $\pm$ 11\\
       & 7229, 40  &  27 $\pm$ 14  &  29 $\pm$ 15\\
\end{tabular}
\end{ruledtabular}
\end{table*}

\begin{table*}[b]
\caption{Integrated luminosities required for $3\sigma$ evidence and $5\sigma$ discovery of the $T_{bc}$ state under different baseline background assumptions. The calculation assumes $\sigma(T_{bc}) = 18~\mathrm{nb}$ and $\mathcal{B}(T_{bc} \to B^{-} D^{+}) = 0.5$. Entries marked with ``--'' indicate that the required luminosity exceeds $300~\mathrm{fb}^{-1}$. The quoted uncertainties include both statistical and systematic components.}
\label{tab:lumifor5sigma_18nb}
\begin{ruledtabular}
\begin{tabular}{cccc}
DPS $\sigma_{\mathrm{eff}}$ [mb] & m($T_{bc}$), $\Gamma$($T_{bc}$) [MeV] 
& $\mathcal{L}_{\mathrm{int}}$ [$\mathrm{fb}^{-1}$] (3 $\sigma$) 
& $\mathcal{L}_{\mathrm{int}}$ [$\mathrm{fb}^{-1}$] (5 $\sigma$) \\
\hline
\multirow{4}{*}{5}      & 7167, 0.5 &  117 $\pm$ 65  &   318 $\pm$ 162 \\
       & 7167, 5.0 &  123 $\pm$ 68  &  335 $\pm$ 170 \\
       & 7229, 10  &  290 $\pm$ 152  &  - \\
       & 7229, 40  &  585 $\pm$ 298  & - \\
\hline
\multirow{4}{*}{15}    & 7167, 0.5 &  64 $\pm$ 37  &  175 $\pm$ 90\\
       & 7167, 5.0 &  68 $\pm$ 39  &  183 $\pm$ 94\\
       & 7229, 10  &  124 $\pm$ 69  &  339 $\pm$ 172 \\
       & 7229, 40  &  226 $\pm$ 120  &  - \\
\hline
\multirow{4}{*}{30}     & 7167, 0.5 &  50 $\pm$ 29   &  137 $\pm$ 70\\
       & 7167, 5.0 &  52 $\pm$ 30  &  143 $\pm$ 73\\
       & 7229, 10  &  82 $\pm$ 47  &  223 $\pm$ 114\\
       & 7229, 40  &  134 $\pm$ 74  &  368 $\pm$ 186 \\
\end{tabular}
\end{ruledtabular}
\end{table*}

\subsection{Minimum Cross Section times Branching Fraction for a $5\sigma$ Discovery of $T_{bc}$ at LHCb}

Tables~\ref{tab:SigmaBr_50invfb} and~\ref{tab:SigmaBr_300invfb} show the minimum observable values of the product $\sigma(T_{bc}) \times \mathcal{B}(T_{bc} \to B^- D^+)$ required to achieve a $5\sigma$ discovery of $T_{bc}$ for different integrated-luminosity scenarios. Table~\ref{tab:SigmaBr_50invfb} corresponds to $\mathcal{L}_{\mathrm{int}}=50~\mathrm{fb}^{-1}$, combining data from Runs~2–4, while Table~\ref{tab:SigmaBr_300invfb} corresponds to $\mathcal{L}_{\mathrm{int}}=300~\mathrm{fb}^{-1}$, including data from Runs~2–5. Increasing the integrated luminosity from $50~\mathrm{fb}^{-1}$ to $300~\mathrm{fb}^{-1}$ substantially lowers the minimum observable $\sigma(T_{bc})\times \mathcal{B}(T_{bc} \to B^- D^+)$, thereby enhancing statistical sensitivity. In particular, the minimum observable value is reduced by a factor of three or more, depending on the assumed $T_{bc}$ parameters and background scenarios.

Assuming $\sigma(T_{bc}) \times \mathcal{B}(T_{bc} \to B^- D^+)$ lies in the range $20$–$60~\mathrm{nb}$, a $5\sigma$ observation of $T_{bc}$ could be achieved by the end of Run~4. Smaller values, in the range $5$–$25~\mathrm{nb}$, would require the full Run~5 dataset. These results emphasize that both the integrated luminosity and the production cross section times branching fraction are key determinants of the discovery potential.

\begin{table*}[b]
\caption{Minimum observable value of $\sigma(T_{bc}) \times \mathcal{B}(T_{bc} \to B^{-} D^{+})$ required to achieve a $5\sigma$ discovery of $T_{bc}$ at $\mathcal{L}_{\mathrm{int}} = 50~\mathrm{fb^{-1}}$ (Runs~2--4), for different $T_{bc}$ parameter assumptions, shown for both the baseline and conservative, maximal-background scenarios. The quoted uncertainties include both statistical and systematic contributions.}
\label{tab:SigmaBr_50invfb}
\begin{ruledtabular}
\begin{tabular}{cccc}
DPS $\sigma_{\mathrm{eff}}$ [mb] & m($T_{bc}$), $\Gamma$($T_{bc}$) [MeV] &  $\sigma(T_{bc}) \times \mathcal{B}$ [nb] (NLO SPS 3FNS) &
$\sigma(T_{bc}) \times \mathcal{B}$ [nb] (LO SPS 4FNS, max)\\
\hline
\multirow{4}{*}{5}      & 7167, 0.5 &  30 $\pm$ 11 & 30 $\pm$ 11\\
       & 7167, 5.0 &  31 $\pm$ 11 & 31 $\pm$ 12\\
       & 7229, 10  &  43 $\pm$ 15 & 44 $\pm$ 15\\
       & 7229, 40  &  60  $\pm$ 19 & 60 $\pm$ 19\\
\hline
\multirow{4}{*}{15}     & 7167, 0.5 &  23 $\pm$ 10 & 23 $\pm$ 10\\
       & 7167, 5.0 &  24 $\pm$ 10 & 24 $\pm$ 10\\
       & 7229, 10  &  31 $\pm$ 12 & 31 $\pm$ 12\\
       & 7229, 40  &  40 $\pm$ 14 & 41 $\pm$ 14\\
\hline
\multirow{4}{*}{30}    & 7167, 0.5 &  21 $\pm$ 9 & 21 $\pm$ 9\\
       & 7167, 5.0 &  21 $\pm$ 10 & 22 $\pm$ 10\\
       & 7229, 10  &  26 $\pm$ 11 & 27 $\pm$ 11\\
       & 7229, 40  &  32 $\pm$ 12 & 34 $\pm$ 13\\
\end{tabular}
\end{ruledtabular}
\end{table*}

\begin{table*}[b]
\caption{Minimum observable $\sigma(T_{bc}) \times \mathcal{B}(T_{bc} \to B^{-} D^{+})$ required for a $5\sigma$ discovery of $T_{bc}$ at $\mathcal{L}_{\mathrm{int}} = 300~\mathrm{fb^{-1}}$ (combined Run~2, Run~3, and Run~4), under various $T_{bc}$ parameter assumptions, is shown for both the baseline and conservative maximal-background scenarios. The quoted uncertainties include both statistical and systematic contributions.}
\label{tab:SigmaBr_300invfb}
\begin{ruledtabular}
\begin{tabular}{cccc}
DPS $\sigma_{\mathrm{eff}}$ [mb] & m($T_{bc}$), $\Gamma$($T_{bc}$) [MeV] &  $\sigma(T_{bc}) \times \mathcal{B}$ [nb] (NLO SPS 3FNS) &
$\sigma(T_{bc}) \times \mathcal{B}$ [nb] (LO SPS 4FNS, max)\\
\hline
\multirow{4}{*}{5}      & 7167, 0.5 &  9 $\pm$ 3  & 10 $\pm$ 3\\
       & 7167, 5.0 &  10 $\pm$ 3 & 10 $\pm$ 3\\
       & 7229, 10  &  15 $\pm$ 4 & 15 $\pm$ 4\\
       & 7229, 40  &  21 $\pm$ 6 & 22 $\pm$ 6\\
\hline
\multirow{4}{*}{15}     & 7167, 0.5 &  6 $\pm$ 2  & 6 $\pm$ 2\\
       & 7167, 5.0 &  6 $\pm$ 2  & 7 $\pm$ 2\\
       & 7229, 10  &  10 $\pm$ 3 & 10 $\pm$ 3\\
       & 7229, 40  &  13 $\pm$ 4 & 14 $\pm$ 4\\
\hline
\multirow{4}{*}{30}     & 7167, 0.5 &  5 $\pm$ 2  & 5 $\pm$ 2\\
       & 7167, 5.0 &  5 $\pm$ 2  & 6 $\pm$ 2\\
       & 7229, 10  &  7 $\pm$ 2  & 8 $\pm$ 3\\
       & 7229, 40  &  10 $\pm$ 3 & 11 $\pm$ 3 \\
\end{tabular}
\end{ruledtabular}
\end{table*}

\section{SUMMARY}

In this study, we investigate the discovery potential of the $T_{bc}$ state at the LHCb experiment using a phenomenological approach. The background is modeled with \mgshort\ and \pythia\ generators, with simulations tuned to published LHCb measurements. We scan the parameter space of the $T_{bc}$ mass (7167–7229 MeV), width (0.5–40 MeV), production cross section ($\sigma(T_{bc}) = 0.3, 18, 103~\mathrm{nb}$), and effective DPS cross section ($\sigma_{\mathrm{eff}} = 5, 15, 30~\mathrm{mb}$) to determine the integrated luminosity required for a $5\sigma$ discovery. 

Our results indicate that, for the optimistic production cross section of $\sigma(T_{bc}) = 103~\mathrm{nb}$ and a branching fraction of $\mathcal{B}(T_{bc}\to B^-D^+)=0.5$, a $5\sigma$ discovery is expected by the end of Run~4, except in the case of a heavy, broad $T_{bc}$ state ($m_{T_{bc}}=7229$ MeV, $\Gamma(T_{bc})=40$ MeV) with a large DPS background ($\sigma_{\mathrm{eff}}=5$~mb), for which the significance is nevertheless still sizable. 
For the intermediate estimate of $\sigma(T_{bc}) = 18~\mathrm{nb}$, most parameter choices yield a $3\sigma$ evidence with the full Run~5 dataset, while only the most favorable scenarios reach $5\sigma$. In the most conservative scenario of $\sigma(T_{bc}) = 0.5~\mathrm{nb}$, no signal would be observable even with $300~\mathrm{fb}^{-1}$.

We also evaluated the minimum observable $\sigma(T_{bc}) \times \mathcal{B}(T_{bc} \to B^{-} D^{+})$ required for a $5\sigma$ discovery under different integrated luminosity scenarios. With $50~\mathrm{fb}^{-1}$ of data (Run~2 to Run~4), a discovery is expected if $\sigma(T_{bc}) \times \mathcal{B}(T_{bc} \to B^- D^+)$ lies between $20~\mathrm{nb}$ and $60~\mathrm{nb}$. For smaller cross sections in the range $5~\mathrm{nb}$ to $25~\mathrm{nb}$, the full Run~5 dataset ($300~\mathrm{fb}^{-1}$) is needed. 

In conclusion, this work provides a quantitative, data-driven assessment of the $T_{bc}$ discovery potential at LHCb, addressing the challenge of background modeling for prompt $\bar{B}D$ production. A systematic search for the $T_{bc}$ at LHCb is feasible, with promising prospects for evidence ($3\sigma$) and potential discovery ($5\sigma$), depending on the integrated luminosity and $T_{bc}$ production cross section. Even in the absence of a signal, these results provide valuable theoretical input by setting upper limits on $\sigma(T_{bc}) \times \mathcal{B}(T_{bc} \to B^-D^+)$, thereby offering experimental constraints to guide hadron models. Together, these findings establish a foundation for further exploration of exotic hadrons at LHCb.

\acknowledgments
We are grateful to Xiaojie Jiang, Ivan Polyakov and Xuhao Yuan for useful discussions. Views and opinions expressed are however, those of the authors only and do not necessarily reflect those of the European Union or the European Research Council Executive Agency. Neither the European Union nor the granting authority can be held responsible for them.

\bibliographystyle{cpc}

\bibliography{reference}

\end{document}